\documentclass[11pt, a4paper]{article}
\pdfoutput=1

\usepackage{jheppub}

\usepackage[T1]{fontenc} %Permite escribir acentos: á en lugar de \'a etc.
\usepackage{physics} % ftp://ftp.dante.de/tex-archive/macros/latex/contrib/physics/physics.pdf
\usepackage{booktabs} % Nice tables
\usepackage{multirow} % Nonstandard tables
\usepackage[bottom]{footmisc}

\usepackage{tikz}
\usetikzlibrary{arrows,shapes}
\usetikzlibrary{trees}
\usetikzlibrary{matrix,arrows} 				% For commutative diagram
\usetikzlibrary{positioning}				% For "above of=" commands
\usetikzlibrary{calc,through}				% For coordinates
\usetikzlibrary{decorations.pathreplacing}  % For curly braces
\usepackage{pgffor}							% For repeating patterns
\usetikzlibrary{decorations.pathmorphing}	% For Feynman Diagrams
\usetikzlibrary{decorations.markings}
\tikzset{
	vector/.style={decorate, decoration={snake}, draw},
	provector/.style={decorate, decoration={snake,amplitude=2.5pt}, draw},
	antivector/.style={decorate, decoration={snake,amplitude=-2.5pt}, draw},
	fermion/.style={draw=black, postaction={decorate},
	decoration={markings,mark=at position .55 with {\arrow[draw=black]{>}}}},
	fermionbar/.style={draw=black, postaction={decorate},
	decoration={markings,mark=at position .55 with {\arrow[draw=black]{<}}}},
	fermionnoarrow/.style={draw=black},
	gluon/.style={decorate, draw=black,
	decoration={coil,amplitude=4pt, segment length=5pt}},
	scalar/.style={dashed,draw=black, postaction={decorate},
	decoration={markings,mark=at position .55 with {\arrow[draw=black]{>}}}},
	scalarbar/.style={dashed,draw=black, postaction={decorate},
	decoration={markings,mark=at position .55 with {\arrow[draw=black]{<}}}},
	scalarnoarrow/.style={dashed,draw=black},
	electron/.style={draw=black, postaction={decorate},
	decoration={markings,mark=at position .55 with {\arrow[draw=black]{>}}}},
	bigvector/.style={decorate, decoration={snake,amplitude=4pt}, draw},
}

\def\tr{\mathrm{tr}}

\def\dm{\Delta m^2}
\def\mo{\tilde m^2_0}
\def\mobar{\tilde {\bar m}^2_0}
\def\dmt{\Delta \tilde m^2}
\def\dmtbar{\Delta \tilde {\bar m}^2}

\newcommand{\asym}[1]{\mathcal{A}^\mathrm{#1}}

\newcommand{\f}[1]{f_\mathrm{#1}}

\title{Signatures of the genuine and matter-induced components of the CP violation asymmetry in neutrino oscillations}

\author{Jos\'e Bernab\'eu}
\author{and Alejandro Segarra}
\affiliation{Departament de F\'isica Te\`orica and IFIC, Universitat de Val\`encia - CSIC, E-46100, Spain}

\emailAdd{Jose.Bernabeu@uv.es}
\emailAdd{Alejandro.Segarra@uv.es}

\abstract{
	CP asymmetries for neutrino oscillations in matter can be disentangled into 
	the matter-induced CPT-odd (T-invariant) component and 
	the genuine T-odd (CPT-invariant) component. 
	For their understanding in terms of the relevant ingredients, 
	we develop a new perturbative expansion in both $\dm_{21},\, \abs{a} \ll\abs{\dm_{31}}$
	without any assumptions between $\dm_{21}$ and $a$,
	and study the subtleties of the vacuum limit in the two terms of the CP asymmetry, 
	moving from the CPT-invariant vacuum limit $a \to 0$ 
	to the T-invariant limit $\dm_{21} \to 0$. 
	In the experimental region of terrestrial accelerator neutrinos, 
	we calculate their approximate expressions from which we prove that,
	at medium baselines, the CPT-odd component is small and nearly
	$\delta$-independent, so it can be subtracted from the experimental CP asymmetry 
	as a theoretical background, provided the hierarchy is known.
	At long baselines, on the other hand, we find that
	(i) a Hierarchy-odd term in the CPT-odd component 
	dominates the CP asymmetry for energies above the first oscillation node, 
	and (ii) the CPT-odd term vanishes, independent of the CP phase $\delta$, at 
	$E =0.92~\mathrm{GeV}\,(L/1300~\mathrm{km})$
	near the second oscillation maximum, 
	where the T-odd term is almost maximal and proportional to $\sin\delta$. 
	A measurement of the CP asymmetry in these energy regions 
	would thus provide separate information on
	(i) the neutrino mass ordering, 
	and (ii) direct evidence of genuine CP violation in the lepton sector.
}

\begin{document}

\maketitle
\flushbottom

\section{Introduction}

One of the most important open questions in fundamental physics is the
existence of CP symmetry breaking in the lepton sector. A positive answer could
open the door to understand the matter-antimatter asymmetry of the Universe
through Leptogenesis~\cite{baryogenesis} at higher energy scales. 
Next generation neutrino flavor oscillation experiments, like T2HK~\cite{HK} and DUNE~\cite{DUNE}, 
have this challenge as their priority aim. But the
neutrino propagation from terrestrial accelerator facilities is not taking
place in vacuum and there are matter effects~\cite{MSW-W,MSW-MS}
induced by the CP-asymmetric
interaction with Earth. In the quest for a direct evidence of CP violation in
the lepton sector, we have recently derived a theorem~\cite{disentanglingPRL} for 
the observable CP asymmetry
in neutrino oscillations propagating in matter 
\begin{equation}
	\label{eq:ACP}
	\asym{CP}_{\alpha\beta} \equiv P_{\alpha\beta}(\nu)-P_{\alpha\beta}(\bar\nu) = 
	\asym{CPT}_{\alpha\beta} + \asym{T}_{\alpha\beta}\,,
\end{equation}
where $P_{\alpha\beta}$ is the probability for any flavor transition $\alpha \to \beta$.
Its achievement consists in providing the disentanglement of the genuine $\asym{T}_{\alpha\beta}$
and matter-induced $\asym{CPT}_{\alpha\beta}$ components of CP-violating (CPV) 
$\asym{CP}_{\alpha\beta}$
based on the concept that 
they have different properties under the other discrete symmetries Time
Reversal T and CPT:
$\asym{CPT}_{\alpha\beta}$ is CPT-odd T-invariant, whereas
$\asym{T}_{\alpha\beta}$ is T-odd CPT-invariant.
These two components are distinctly identified 
by their behavior as functions of the baseline $L$:
whereas the matter-induced component $\asym{CPT}_{\alpha\beta}$ is an even function of $L$, 
the genuine component $\asym{T}_{\alpha\beta}$ is odd in $L$. 
For a description of the effective Hamiltonian in which the matter-induced term 
is generated by the parameter $a = 2 E V$ of the matter potential $V$ 
due to charged current interactions between 
electron neutrinos and electrons and the genuine CPV term is generated by a phase $\delta$ in the
three-family PMNS mixing matrix, the $\asym{CPT}_{\alpha\beta}$ and $\asym{T}_{\alpha\beta}$ 
components of the CP asymmetry have
the correct behavior, for any neutrino energy $E$ and baseline $L$, in
these parameters: 
whereas $\asym{CPT}_{\alpha\beta}$ is odd in $a$ $\forall \delta$,
$\asym{T}_{\alpha\beta}$ is odd in $\sin\delta$ $\forall a$. 
In addition, the change from a Normal Hierarchy in the ordering 
of the neutrino mass spectrum to an Inverted Hierarchy leads to 
the result that 
$\asym{T}_{\alpha\beta}$ remains invariant and,
for energies above the first oscillation node,
$\asym{CPT}_{\alpha\beta}$ changes its sign.

The planned experiments T2HK and DUNE consider the golden transition
$\nu_\mu \xrightarrow{\;L\;} \nu_e$
for neutrinos propagating in the Earth mantle with fixed $L$
and a continuum energy spectrum $E$. In the search of interesting
experimental signatures able to separate the $\asym{CPT}_{\mu e}$ and $\asym{T}_{\mu e}$ components
of $\asym{CP}_{\mu e}$, we study in this paper the characteristic energy dependencies
of the two terms. In doing so, we develop appropriate analytical relations between 
the quantities in matter and the quantities in vacuum 
in order to provide guiding paths in our scrutiny of the peculiar
behavior of the separate $\asym{CPT}_{\mu e}$ and $\asym{T}_{\mu e}$ components with energy. 

The paper is organized as follows. 
Section~\ref{sec:theorem} identifies the two components, 
matter-induced $\asym{CPT}_{\alpha\beta}$ and genuine $\asym{T}_{\alpha\beta}$, 
of the CP asymmetry $\asym{CP}_{\alpha\beta}$ by means of rephasing-invariant mixings 
and neutrino masses in matter.
The conceptual basis of the Disentanglement Theorem is provided
by the different behavior under T and CPT symmetries.
In Section~\ref{sec:desarrollos} we build a consistent perturbative expansion 
of the relevant ingredients in terms of the vacuum parameters 
with $\dm_{21}\ll\abs{\dm_{31}}$ and $\abs{U_{e3}}^2\ll 1$, 
with the interaction parameter $a$ moving from above to below $\dm_{21}$. 
This approach will allow us to understand 
the intricacies of the ordering of the two limits
$a \to 0$ and $\dm_{21} \to 0$.
In Section~\ref{sec:AT} we check that our analytic expansions are an
excellent approximation to the exact unique rephasing-invariant CP-odd mixing 
appearing in the genuine $\asym{T}_{\mu e}$ component of the CP asymmetry,
leading in fact to $\asym{T}_{\mu e}$ as in vacuum
by an interesting compensation of matter effects between 
the mixing and oscillation factors.
In Section~\ref{sec:Hier} we discuss the dependence of these observables on the
neutrino mass hierarchy, 
proving that it is determined by the sign of 
$\asym{CPT}_{\mu e}$ at high energies,
whereas $\asym{T}_{\mu e}$ is blind to it.
Section~\ref{sec:signatures} makes a scan of the energy dependencies 
of the two $\asym{CPT}_{\mu e}$ and $\asym{T}_{\mu e}$ components,
identifying zeros and extremal values of these functions. 
The occurrence of magic energies,
in which $\asym{CPT}_{\mu e}$ vanishes ---independent of $\delta$--- 
and $\abs{\asym{T}_{\mu e}}$ is maximal, will be understood. 
In Section~\ref{sec:conclusions} we discuss our conclusions and outlook.

\section{The CP Asymmetry Disentanglement Theorem}
\label{sec:theorem}
Neutrino oscillations in matter are described through the effective Hamiltonian 
in the flavor basis~\cite{MSW-W,Matter-Hamiltonian-Barger, Matter-Hamiltonian-Kuo, Matter-Hamiltonian-Zaglauer, Matter-Hamiltonian-Krastev, Matter-Hamiltonian-Parke}
\begin{equation}
	\label{eq:Hf}
	H =
	\frac{1}{2E}
	\left\{
		U
	\mqty[
		m_1^2 &0 &0\\
		0 &m_2^2 &0\\
		0 &0 &m_3^2
	] 
	U^\dagger
	+
	\mqty[
		\,a\, &\,0\, &\,0\,\\
		0 &0 &0\\
		0 &0 &0
	] \right\}
	=\frac{1}{2E}\; \tilde U \tilde M^2 \tilde U^\dagger\,,
\end{equation}
where the first term describes neutrino oscillations in vacuum
and the second one accounts for matter effects. The $a$ parameter 
is given by $a = 2EV$, with $V$ the interaction potential with matter
and $E$ the relativistic neutrino energy. For antineutrinos, $U\to U^*$,
originating a genuine CP violation effect through a CP phase $\delta$ in 
$U_\mathrm{PMNS}$, as well as $a \to -a$, originating matter-induced CP violation. In this
description, the genuine CPV observable has to be odd in $\sin\delta$, whereas the
matter-induced CPV effect has to be odd in $a$. 
All neutrino masses $(\tilde M^2)$ and mixings $(\tilde U)$ in matter,
i.e. eigenvalues and eigenstates of $H$, 
can be calculated in terms of the parameters in the vacuum Hamiltonian 
$(M^2,\, U)$ and $a$, 
as studied in Section~\ref{sec:desarrollos}.

The exact Hamiltonian leads to the flavor oscillation probabilities
for any $\alpha \to \beta$ transition
\begin{equation}
	\label{eq:Pab}
	P(\nu_\alpha \to \nu_\beta)
	= \delta_{\alpha\beta}
	-4\sum_{j<i}\mathrm{Re}~\tilde J_{\alpha\beta}^{ij}\,
	\sin^2 \tilde \Delta_{ij}\,
	-2\sum_{j<i}\mathrm{Im}~\tilde J_{\alpha\beta}^{ij}\,
	\sin 2\tilde \Delta_{ij}\,,
\end{equation}
where 
$\tilde J_{\alpha\beta}^{ij} \equiv 
\tilde U_{\alpha i} \tilde U^*_{\alpha j}
\tilde U^*_{\beta i} \tilde U_{\beta j}$
are the rephasing-invariant mixings
and
$ \tilde \Delta_{ij} \equiv \frac{\Delta\tilde m^2_{ij} L}{4 E}$.
Notice that both 
$\tilde J_{\alpha\beta}^{ij}$
and
$\dmt_{ij} $
are energy dependent in matter.
Antineutrino oscillations are given in general by the same expression
with different masses $(\Delta {\tilde {\bar m}}^2_{ij} )$ 
and mixings $(\tilde {\bar J}_{\alpha\beta}^{ij})$,
so one can explicitly write the CP asymmetry $\asym{CP}_{\alpha\beta}$
defined in Eq.~(\ref{eq:ACP}).

If CPT holds, as assumed in vacuum, necessarily
$\Delta {{\bar m}}^2_{ij} = \Delta {{ m}}^2_{ij}$
and $ {\bar J}_{\alpha\beta}^{ij} = ({J}_{\alpha\beta}^{ij})^* $.
Therefore, all $L$-even terms will cancel out in $\asym{CP}_{\alpha\beta}$,
proving they are CPT-violating.
On the other hand, the absence of genuine CP violation leads to
real $\tilde J^{ij}_{\alpha\beta}$
and $\tilde {\bar J}^{ij}_{\alpha\beta}$,
so all transition probabilities $P_{\alpha\beta}$ are $L$-even functions.
This result shows that $L$-odd terms in $\asym{CP}_{\alpha\beta}$ are T-violating.

From the different behavior 
of each of these terms %in Eq.~(\ref{eq:Pab})
under the discrete T and CPT symmetry transformations,
one derives the \textbf{Asymmetry Disentanglement Theorem}~\cite{disentanglingPRL}
by separating the observable CP asymmetry 
in any flavor transition into $L$-even (CPT-violating) and $L$-odd (T-violating) functions,
$ \mathcal{A}^\mathrm{CP}_{\alpha\beta} = 
\mathcal{A}^\mathrm{CPT}_{\alpha\beta} + 
\mathcal{A}^\mathrm{T}_{\alpha\beta}$,
\begin{subequations}
	\label{eqs:theorem_asyms}
	\begin{align}
		\mathcal{A}^\mathrm{CPT}_{\alpha\beta} &=
			-4 \sum\limits_{j<i} \left[ 
				\mathrm{Re}~\tilde J_{\alpha\beta}^{ij}\, \sin^2 \tilde \Delta_{ij}
				-
				\mathrm{Re}~\tilde {\bar J}_{\alpha\beta}^{ij}\, \sin^2 \tilde {\bar \Delta}_{ij}
			\right]\,,\\
		\mathcal{A}^\mathrm{T}_{\alpha\beta} &=
			-2\sum\limits_{j<i} \left[ 
				\mathrm{Im}~\tilde J_{\alpha\beta}^{ij}\, \sin 2\tilde \Delta_{ij}
				-
				\mathrm{Im}~\tilde {\bar J}_{\alpha\beta}^{ij}\, \sin 2\tilde {\bar \Delta}_{ij}
			\right]\,.
	\end{align}
\end{subequations}
Let us emphasize that
not only $\asym{CPT}_{\alpha\beta}$ is CPT-violating 
and $\asym{T}_{\alpha\beta}$ is T-violating,
we also find that $\asym{CPT}_{\alpha\beta}$ is T-invariant 
and $\asym{T}_{\alpha\beta}$ is CPT-invariant. 
In this sense the two terms are truly disentangled.
To prove these properties, we analyze both CPT and T transformations.

Under CPT: $\{ \Delta {\tilde{m}}^2_{ij} \leftrightarrow \Delta {\tilde{\bar m}}^2_{ij},\,
{\tilde J}_{\alpha\beta}^{ij} \leftrightarrow (\tilde{\bar J}_{\alpha\beta}^{ij})^* \}$,
neutrino and antineutrino terms in $\asym{CPT}_{\alpha\beta}$ are interchanged, 
so $\asym{CPT}_{\alpha\beta}$ changes its sign.
This sign in $\asym{T}_{\alpha\beta}$ is compensated by the change of sign in both
Im$\,{\tilde J}_{\alpha\beta}^{ij}$ and Im$\,\tilde{\bar J}_{\alpha\beta}^{ij} $,
leaving $\asym{T}_{\alpha\beta}$ invariant.
Under T: $\{  \tilde J_{\alpha\beta}^{ij} \to (\tilde J_{\alpha\beta}^{ij})^*,\,
\tilde {\bar J}_{\alpha\beta}^{ij} \to (\tilde {\bar J}_{\alpha\beta}^{ij})^*\}$,
the only change in the asymmetries is a change of sign in all imaginary parts,
changing the sign of $\asym{T}_{\alpha\beta}$ 
and leaving $\asym{CPT}_{\alpha\beta}$ invariant.
	
These properties lead cleanly to the disentanglement of
$ \mathcal{A}^\mathrm{CP}_{\alpha\beta} = 
\mathcal{A}^\mathrm{CPT}_{\alpha\beta} + 
\mathcal{A}^\mathrm{T}_{\alpha\beta}$,
where 
$\mathcal{A}^\mathrm{CPT}_{\alpha\beta}$
is T-invariant (even in $\sin\delta$) and CPT-odd in $a$, 
whereas 
$\mathcal{A}^\mathrm{T}_{\alpha\beta}$ 
is CPT-invariant (even in $a$) and T-odd in $\sin\delta$.
As a consequence, 
$\mathcal{A}^\mathrm{CPT}_{\alpha\beta}$
vanishes for $a = 0$ $\forall \delta$ and 
$\mathcal{A}^\mathrm{T}_{\alpha\beta}$
vanishes for $\delta = 0,\, \pi$ $\forall a$. 
These complementary behaviors of the
%two components of the observable CP asymmetry will be illustrated 
%both analytically and numerically for a given pair of $(E,\, L)$ values.
two components of the experimental CP asymmetry identify 
the CPV component $\mathcal{A}^\mathrm{T}_{\alpha\beta}$ as CPT-invariant and thus a fully genuine CPV observable, whereas
the CPV component $\mathcal{A}^\mathrm{CPT}_{\alpha\beta}$ is T-invariant and thus a fully fake CPV observable.

\section{Analytic perturbation expansions}
\label{sec:desarrollos}

To understand the behavior of $\asym{CPT}_{\alpha\beta}$ and $\asym{T}_{\alpha\beta}$ 
in Eqs.~(\ref{eqs:theorem_asyms}) required by the CPT and T symmetries,
as proved in the previous Section,
we proceed to their analytic study for neutrino oscillations in matter
of constant density.
Notice that the formal description of the system is equivalent to
neutrino oscillations in vacuum, if one parametrizes matter effects as a
redefinition of neutrino masses and mixings. However, this redefinition is 
strongly dependent on the neutrino energy, so it does not provide a clear insight
into the intrinsic properties of the system. 
A useful description should write all observables in matter,
relevant to our two components of the experimentally accessible CP asymmetry,
as functions of the vacuum parameters and the matter potential $a$.
A similar methodology is being applied to calculations of the 
T~\cite{petcov} and CPT~\cite{shunCPT}
asymmetries in matter.

The search of these formulae unavoidably finds the same issue:
an exact description of the matter effects in neutrino oscillations leads to
cumbersome expressions which do not provide a clear understanding~\cite{xing}.
The way to simplify the results is to treat perturbatively the small parameters of the system,
namely $\dm_{21}\ll\abs{\dm_{31}}$ and $\abs{U_{e3}}^2\ll 1$. The most important drawback of this
procedure is that, in perturbing in $\dm_{21}$, the implicit relation $\dm_{21}\ll \abs{a}$ is
also assumed, so one should not expect to reproduce the right vacuum limit $a\to 0$
for all matter ingredients.
Even so, this perturbation theory leads to compact and percent-level precise expressions 
for neutrino oscillation probabilities written in terms of vacuum parameters only~\cite{cervera},
as well as more precise 
relations mapping  the mixings in matter to the quantities in
vacuum~\cite{parke,ioannisian}.

We develop a new perturbative expansion in both $\dm_{21},\, \abs{a} \ll\abs{\dm_{31}}$ 
without assumptions between $\dm_{21}$ and $a$,
similar to Ref.~\cite{xing_pert},
oriented to the understanding of masses, mixings and the separate behavior of
$\asym{CPT}_{\alpha\beta}$ and $\asym{T}_{\alpha\beta}$
as functions of the different variables.
In doing so, we can check
from our analytic expressions both the vacuum limit $a\to 0$ and the
T-invariant limit $\dm_{21}\to 0$. 
The expansion in $\abs{a}\ll\abs{\dm_{31}}$ holds for energies below a few GeV
taking into account the definite $a$-parity of each component of the CP asymmetry.
This way, we find the most simple expressions
for $\asym{CPT}_{\mu e}$ and $\asym{T}_{\mu e}$ 
at the energies accessible by accelerator experiments,
which are accurate enough to let the reader clearly understand their behavior.

We emphasize that any desired precision can be achieved using a numerical
computation of the neutrino propagation. Our aim is not finding very precise
expansions, 
but precise enough to identify and understand the distinct characteristic
patterns of the energy behavior of the two components
$\asym{CPT}_{\alpha\beta}$ and $\asym{T}_{\alpha\beta}$,
with the objective of serving as a guide for experimental signatures.

\subsection{The crucial role of the reference $\mo$ in matter}

Since a diagonal $m_1^2\, \mathbf{1}$ in the Hamiltonian $H$ in Eq.~(\ref{eq:Hf})
leads to a global phase in time evolution, which is unobservable,
the equivalent Hamiltonian 
$2E\, \Delta H \equiv \Delta H' = U\,\Delta M^2\, U^\dagger +a\, P_e$,
where $\Delta M^2 = \mathrm{diag}(0,\, \dm_{21},\, \dm_{31})$ and
$P_e = \mathrm{diag}(1, 0, 0)$ is the $e$-flavor projector,
is widely used.

Analogously, one could argue that $\tilde m_1^2$ is unobservable in neutrino
oscillations in matter. Even though this is true, one must take into account
that either $m_1^2$ or $\tilde m_1^2$ can be chosen as origin of phases,
but not both of them at the same time
when connecting the parameters in matter to those in vacuum. 
Indeed, one can easily check that
the Hamiltonian $\Delta H'$ has three non-vanishing eigenvalues,
so choosing $m_1^2 = 0$ automatically leads to all $\tilde m_i \neq 0$,
despite one of them being unobservable.

On the following, we call $\mo$ the mass squared in matter leading to 
the relative phase shift between the unobservable global
phases in vacuum and matter, writing the Hamiltonian as 
$\Delta H' = U\,\Delta M^2\, U^\dagger +a\, P_e = 
\mo\, \mathbf{1} + \tilde U\,\Delta {\tilde M}^2\, {\tilde U}^\dagger$.
In this notation, the three eigenvalues of $\Delta H'$ will be the
reference scale in matter $\mo$ and the two observable 
mass squared differences in matter $\dmt_{ij}$.

As proposed in Ref.~\cite{xing}, we choose to diagonalize the Hamiltonian in
the vacuum eigenbasis, 
$\Delta H'_{ij} = \Delta M^2 + a\, U^\dagger P_e U = 
\mo \mathbf{1} + V \Delta \tilde M^2 V^\dagger$.
Since the eigenvalues are basis independent and real, this cleanly shows that 
$\mo$ and $\dmt_{ij}$ can only depend on $(\dm_{ij},\, \abs{U_{ei}},\, a)$.
Moreover, this factorizes the mixing matrix in matter into $\tilde U = UV$,
where $U$ is the (vacuum) PMNS matrix and $V$ is the change of basis between
vacuum and matter eigenstates, which must go to the identity when $a\to 0$.

The Hamiltonian being a $3\times 3$ matrix leads to the characteristic equation
\begin{equation}
	\label{eq:CharEq}
	p(\lambda) \equiv
	- \lambda^3 
	+ \lambda^2\, \tr(\Delta H')
	+ \frac{1}{2}\,\lambda \left[ \tr[(\Delta H')^2] - \tr^2(\Delta H') \right]
	+ \det(\Delta H')
	= 0\,,
\end{equation}
where $p(\lambda)$ is the characteristic polynomial of $\Delta H'$ and 
its roots provide the three neutrino squared masses in matter $\lambda_i$, 
and the observable $\dmt_{ij}=\lambda_i -\lambda_j$.
The three invariants can be easily calculated,
\begin{subequations}
	\begin{align}
		\tr(\Delta H') &= \dm_{21} + \dm_{31} + a\,,\\
		\tr[(\Delta H')^2] &= (\dm_{21})^2 + (\dm_{31})^2 + a^2 
		+ 2a \left[ \dm_{21} \abs{U_{e2}}^2 + \dm_{31} \abs{U_{e3}}^2 \right]\,, \\
		\det(\Delta H') &= a\abs{U_{e1}}^2\dm_{21}\dm_{31}\,.
	\end{align}
\end{subequations}

\begin{figure}[t!]
	\centering
	\begin{tikzpicture}[line width=1 pt, scale=1.5]
		\node at (0,0){
			\includegraphics[width=0.475\textwidth]{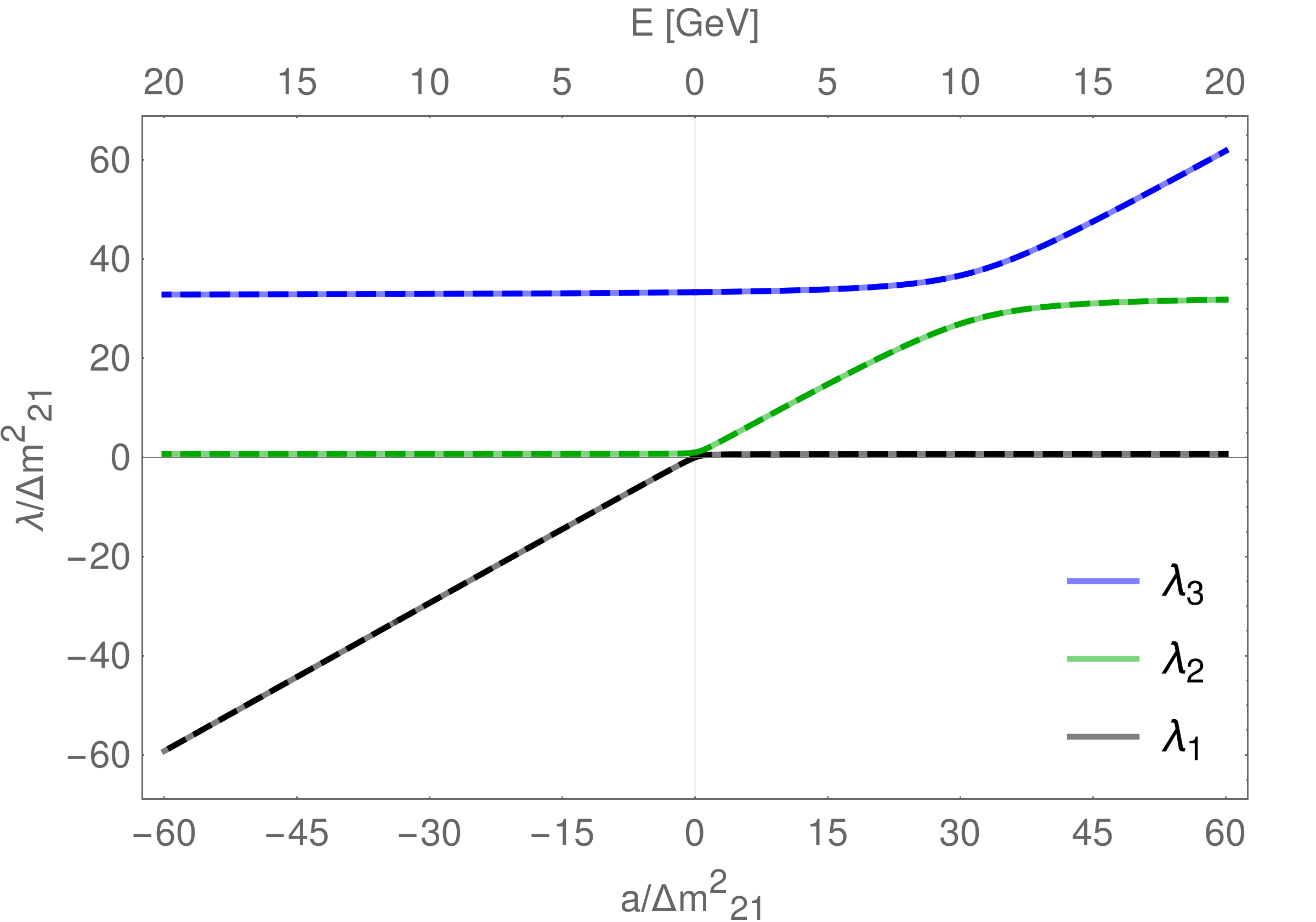}
		};

		\node at (1.45,-1.2)[above left]{NH};
	\end{tikzpicture}
	\hfill
	\begin{tikzpicture}[line width=1 pt, scale=1.5]
		\node at (0,0){
			\includegraphics[width=0.475\textwidth]{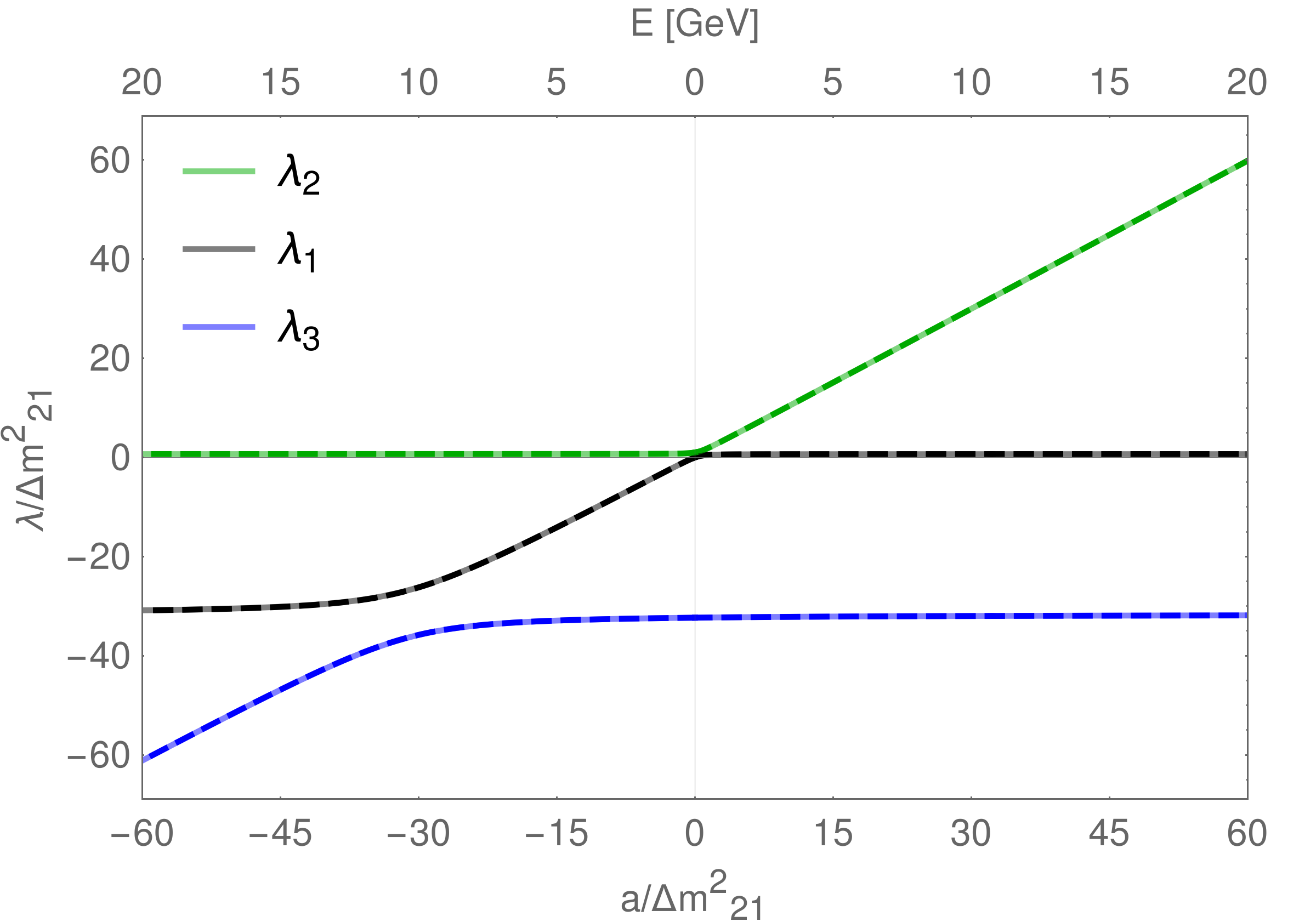}
		};

		\node at (-1.2,1.25)[below right]{IH};
	\end{tikzpicture}
	\caption{Eigenvalues $\lambda_i$ of the mass matrix in matter
	in units of $\dm_{21}$, for both neutrinos $(a>0)$ and
	antineutrinos $(a < 0)$. 
	The horizontal axis shows both the evolution of the matter parameter $a$ at
	fixed energy (lower labels), i.e. changing the matter density,
	and as function of the energy if the constant
	density is chosen as that of the Earth mantle (upper labels).
	Both the exact (dashed) and the analytical (solid)
	results from Eqs.~(\ref{eqs:foralla}) are shown 
	to illustrate the excellence of the analytic approximation.
	Normal Hierarchy ($\lambda_1 < \lambda_2 < \lambda_3$) in the left pannel,
	Inverted Hierarchy ($\lambda_3 < \lambda_1 < \lambda_2$) in the right pannel.}
	\label{fig:mtildes}
\end{figure}

From this straightforward setup of the problem 
we find a fundamental result. Since
\begin{subequations}
	\label{eqs:m0bound}
	\begin{align}
		\label{eqs:m0bound_0}
		p(0) &= a \abs{U_{e1}}^2 \dm_{21} \dm_{31} \geq 0 \,,\\
		\label{eqs:m0bound_dm21}
		p(\dm_{21}) &= a \abs{U_{e2}}^2 \dm_{21} (\dm_{21}-\dm_{31}) \leq 0,
	\end{align}
\end{subequations}
at least one of the eigenvalues of $\Delta H'$ will always lie in the range
$[0,\, \dm_{21}]$. All $\dmt_{ij}$ are known to be nonbound by $\dm_{21}$,
as will be shown in Fig.~\ref{fig:mtildes}, 
so we find that $0 \leq \mo \leq \dm_{21}$. 
Although the inequalities in (\ref{eqs:m0bound_0}) and (\ref{eqs:m0bound_dm21})
have been written for Normal
Hierarchy neutrinos, the reader may check that the argument is also valid for
the Inverted Hierarchy and antineutrinos.

Given that physically $\dm_{21} \ll\abs{\dm_{31}}$, 
this result shows that $\mo$ is also a good perturbative parameter.
Therefore, we focus in this Subsection on writing the 
two observable $\dmt_{ij}$ exactly as functions of
$(\dm_{ij}, \mo, \abs{U_{ei}}, a)$, which gives enough information
to calculate all observables of neutrino oscillations in matter.

A simple way to calculate the physical $\dmt_{ij}$ is the diagonalization of the displaced Hamiltonian
$\Delta H'_{ij} -\mo \mathbf{1} = \Delta M^2 + a\, U^\dagger P_e U -\mo \mathbf{1}
= V \Delta \tilde M^2 V^\dagger$. By construction, one of its eigenvalues is zero,
so $\det(\Delta H' -\mo \mathbf{1}) = 0$ and its two non-vanishing eigenvalues 
are given by a quadratic equation,
\begin{align}
	\label{eq:dmt}
	\dmt_\pm &=
	\frac{1}{2} \left( \dm_{21}+\dm_{31}+a -3\mo \right) \pm
		\frac{1}{2}\sqrt{l^2+2\mo (\dm_{21}+\dm_{31}+a)-3(\mo)^2}\,,\\
	%\nonumber
	%&l^2 = \left( \dm_{31}-\dm_{21} \right)^2 + a^2 
	%- 2a\dm_{21} \left( 1-2\abs{U_{e2}}^2 \right)
	%- 2a\dm_{31} \left( 1-2\abs{U_{e3}}^2 \right)\,.\\
	%\nonumber
	%&l^2 = \left( \dm_{31}-a \right)^2 + (\dm_{21})^2 
	%- 4a\dm_{31} \abs{U_{e3}}^2 -2\dm_{21}\dm_{31}
	%- 2a\dm_{21} \left( 1-2\abs{U_{e2}}^2 \right)\,.\\
	\nonumber
	&l^2 \equiv \left( \dm_{31}+\dm_{21}-a \right)^2
	-4\dm_{21}\dm_{31}
	+ 4a\dm_{21} \abs{U_{e2}}^2
	+ 4a\dm_{31} \abs{U_{e3}}^2\,.
\end{align}
From this definition it is clear that $\dmt_+ > \dm_-$, but notice that
$\abs{\dmt_+} > \abs{\dmt_-}$ for Normal Hierarchy, whereas
$\abs{\dmt_-} > \abs{\dmt_+}$ for Inverted Hierarchy.
This expression for $\dmt_\pm$ is a good starting point from which one
can easily derive approximate formulae in the limit 
$\mo~\leq~\dm_{21}~\ll~\abs{\dm_{31}}$. In order to write $\dmt_\pm$ as functions
of vacuum parameters only, this same limit can be used directly in Eq.~(\ref{eq:CharEq})
to find $\mo$ perturbatively, as we will do in the following Subsection.

We finish this Subsection writing explicitly the eigenstates of $\Delta H'_{ij}$
in the canonical basis of mass eigenstates in vacuum,
\begin{equation}
	\label{eq:nutildegeneral}
	\ket{\tilde \nu_i} =
	\frac{1}{N_i}
	\mqty[ 
	a \left[ \left( \lambda_i -\dm_{31} \right) \abs{U_{e2}}^2 +
		\left( \lambda_i - \dm_{21} \right) \abs{U_{e3}}^2 \right] U_{e1}^*	\\ 
		\left[ \lambda_i -a\abs{U_{e1}}^2 \right] \left[ \lambda_i -\dm_{31} \right] U_{e2}^* \\
		\left[ \lambda_i -a\abs{U_{e1}}^2 \right] \left[ \lambda_i -\dm_{21} \right] U_{e3}^*
	]\,,
\end{equation}
where
the normalization factor $N_i$ is needed to ensure
$\braket{\tilde \nu_i}{\tilde \nu_i} = 1$,
and its phase must be chosen so that
$\lim\limits_{a\to 0} \ket{\tilde \nu_i} = \ket{\nu_i}$.

The eigenvalues $\lambda_i$,
labeled according to 
$\lambda_1 < \lambda_2 < \lambda_3$
($\lambda_3 < \lambda_1 < \lambda_2$)
if the hierarchy is normal (inverted),
are given by $\mo$ and $\dmt_\pm$
as shown in Table~\ref{tab:mtildes}
for neutrinos and antineutrinos.
The reason why $\lambda_1 = \mo$ whereas $\bar \lambda_2 = \mobar$
will be explained analytically when exploring the vacuum limit.
This fact is also shown in Fig.~\ref{fig:mtildes} by a numerical determination of the
evolution of the eigenvalues of $\Delta H'$ with the matter parameter $a$. 
All results are produced using the best-fit values
in Ref.~\cite{mariam} for Normal Hierarchy.
To show the theoretical implications of a change in mass hierarchy, 
we compute the Inverted Hierarchy case using 
$\left.\dm_{31}\right|_\mathrm{IH} =- \left.\dm_{32}\right|_\mathrm{NH}$,
which is physically consistent since it keeps the 
absolute value of the largest mass splitting unchanged.

As seen, the eigenvalues that
fulfill $0\leq \mo \leq \dm_{21}$ are $\lambda_1$ and $\bar \lambda_2$,
independently of whether the hierarchy is normal or inverted.
To distinguish these two functions, from now on we will call them
$\lambda_1 \equiv \mo$ and $\bar \lambda_2 \equiv \mobar$.
Notice that, even if both $\mo$ and $\mobar$ are bounded by $\dm_{21}$,
they are necessarily different functions, as seen by their different
vacuum limits 
$\lim\limits_{a\to 0}\mo = 0$
and
$\lim\limits_{a\to 0}\mobar = \dm_{21}$.

\vfill
\begin{table}[t!]
	\caption{Relation between the eigenvalues $\lambda_i$,
	with the convention $\lambda_i \xrightarrow{\;a\to0\;} \dm_{i1}$,
	and the quantities $\mo$ and $\dmt_\pm$ as calculated from Eq.~(\ref{eq:dmt})
	with the corresponding sign of $a$ for $\nu/\bar \nu$
	and the sign and value of $\dm_{31}$ for NH/IH.
	According to hierarchy, the eigenvalues are ordered from larger to smaller. 
	The observable $\dmt_{ij} = \lambda_i - \lambda_j$ can be read from the table.
	}
	\label{tab:mtildes}
	\begin{center}
		\setlength{\tabcolsep}{12pt}
		\begin{tabular}{cll}
			\toprule
			&Neutrinos $(a>0)$
			&Antineutrinos $(a<0)$\\
			\midrule
			NH 
			& $\lambda_3 = \mo + \dmt_+$ 
			& $\bar \lambda_3 = \mobar + \dmt_+$ \\
			$(\dm_{31}>0)$
			& $\lambda_2 = \mo + \dmt_-$ 
			& $\bar \lambda_2 = \mobar$ \\
			& $\lambda_1 = \mo$ 
			& $\bar \lambda_1 = \mobar + \dmt_-$ \\
			\midrule
			IH 
			& $\lambda_2 = \mo + \dmt_+$ 
			& $\bar \lambda_2 = \mobar$ \\
			$(\dm_{31}<0)$
			& $\lambda_1 = \mo$ 
			& $\bar \lambda_1 = \mobar + \dmt_+$ \\
			& $\lambda_3 = \mo + \dmt_-$ 
			& $\bar \lambda_3 = \mobar + \dmt_-$ \\
			\bottomrule
		\end{tabular}
	\end{center}
\end{table}

\subsection{The way to the vacuum limit at fixed $(E,\, L)$}
\label{sec:vacuum_limit}

The perturbation theory used in the literature to make profit 
of the experimental relation $\dm_{21}\ll\abs{\dm_{31}}$ 
also assumes $\dm_{21}\ll \abs{a}$.
In order to ensure that all our expressions will reproduce the right
vacuum limit, which is crucial to study the CPT-invariant limit on
$\asym{CP}_{\alpha\beta}$,
we expand $\dm_{21}\ll\abs{\dm_{31}}$ without any assumption
between $\dm_{21}$ and $a$.
Up to first order in this regime, Eqs.~(\ref{eq:CharEq}) and (\ref{eq:dmt}) reduce to
\begin{subequations}
	\label{eqs:foralla}
	\begin{align}
		\nonumber
		(\mo)^2(\dm_{31}+a)-
		\mo [\dm_{21}\dm_{31}+a\dm_{21}(1-\abs{U_{e2}}^2)+a\dm_{31}(1-\abs{U_{e3}}^2)]\,+
		\hspace{1.55cm}\\
		\label{eq:m0foralla}
		+a \abs{U_{e1}}^2 \dm_{21} \dm_{31} = 0\,,\\
		\nonumber
		\dmt_\pm = \frac{1}{2}(\dm_{31}+a+\dm_{21}-3\mo) \,\pm 
		\hspace{8.5cm}\\
		\pm \frac{1}{2}
		\sqrt{(\dm_{31}-a)^2 + 4\abs{U_{e3}}^2a\dm_{31} 
		-2(\dm_{21}+\mo)(\dm_{31}+a)+4\abs{U_{e2}}^2a\dm_{21}} \,.
	\end{align}
\end{subequations}
These analytical results are shown in Fig.~\ref{fig:mtildes} 
for neutrinos $(a>0)$ and antineutrinos $(a<0)$, 
as well as Normal $(\dm_{31}>0)$ and Inverted $(\dm_{31}<0)$ Hierarchies, 
and they match perfectly the exact numerical values.
Notice that the two solutions of Eq.~(\ref{eq:m0foralla}) in vacuum are $0,\, \dm_{21}$.
From the discussion in the previous Subsection we know that the first one 
corresponds to $\mo$, which is bound by $\dm_{21}$ when $a>0$, 
whereas the second one, $\mobar$, is bound by $\dm_{21}$ when $a<0$.

The appropriate expansion for $\dmt_\pm$ in both $\mo\leq \dm_{21}\ll\abs{\dm_{31}}$
and $\abs{a} \ll\abs{\dm_{31}}$ is
\begin{subequations}
	\label{eqs:dmpm_1st}
	\begin{align}
		\dmt_{+\mathrm{sign}(\dm_{31})} &= \dm_{31} + a \abs{U_{e3}}^2 - \mo\,. \\
		\dmt_{-\mathrm{sign}(\dm_{31})} &= \dm_{21} + a(1-\abs{U_{e3}}^2)-2\mo\,.
	\end{align}
\end{subequations}
The same expressions apply to antineutrinos changing $a\to -a,\; \mo \to \mobar$,
and the dependence on the Hierarchy is implicit in sign$(\dm_{31})$, that
accounts for the interchange of the expressions 
$\dmt_\pm|_\mathrm{NH} \leftrightarrow \dmt_\mp|_\mathrm{IH}$.

The approximation of $\mo$ and $\mobar$, on the other hand, comes from neglecting
$\dm_{31}$-independent terms in Eq.~(\ref{eq:m0foralla}),
\begin{equation}
	\label{eq:m0_6th}
	%(\mo)^2 - [\dm_{21} + a(1-\abs{U_{e3}}^2) ]\mo +a \abs{U_{e1}}^2 \dm_{21}
	%=0\,,
	\mo = \frac{1}{2}\left[\dm_{21}+a(1-\abs{U_{e3}}^2)\right] \pm
	\frac{1}{2}\sqrt{\left[\dm_{21}+a(1-\abs{U_{e3}}^2)\right]^2-
	4\abs{U_{e1}}^2 a \dm_{21}}\,,
\end{equation}
where the $-(+)$ sign corresponds to $\mo\, (\mobar)$.
In order to compare their behavior above and below $\dm_{21}$, 
one can further expand Eq.~(\ref{eq:m0_6th}) in the two regions

\noindent
\begin{minipage}{0.7\textwidth}
	\begin{equation*}
		\mo = \left\{\;
		\begin{aligned}
			&a\abs{U_{e1}}^2
			\left[
				1-\frac{a \abs{U_{e2}}^2}{\dm_{21}}
				+ \cdots
			\right]\,,\\
			& \frac{\dm_{21} \abs{U_{e1}}^2}{1-\abs{U_{e3}}^2}
			\left[
				1-\frac{\dm_{21} \abs{U_{e2}}^2}{a \left(1-\abs{U_{e3}}^2\right)^2}
				+ \cdots
			\right]\,,
		\end{aligned}
		\right.
	\end{equation*}
\end{minipage}
\hfill
\begin{minipage}{0.3\textwidth}
	\begin{subequations}
		\label{eqs:m0_pert}
		\begin{align}
			&\begin{aligned}
				&\mbox{}\\
				&\mbox{}
			\end{aligned}
			a \ll \dm_{21}\\
			&\begin{aligned}
				&\mbox{}\\
				&\mbox{}
			\end{aligned}
			\dm_{21} \ll a
		\end{align}
	\end{subequations}
\end{minipage}

\noindent
\begin{minipage}{0.7\textwidth}
	\begin{equation*}
		\mobar = \left\{\;
		\begin{aligned}
			&\dm_{21} - \abs{a}\abs{U_{e2}}^2
			\left[
				1-\frac{\abs{a} \abs{U_{e1}}^2}{\dm_{21}}
				+ \cdots
			\right]\,,\\
			& \frac{\dm_{21} \abs{U_{e1}}^2}{1-\abs{U_{e3}}^2}
			\left[
				1+\frac{\dm_{21} \abs{U_{e2}}^2}{\abs{a} \left(1-\abs{U_{e3}}^2\right)^2}
				+ \cdots
			\right]\,,
		\end{aligned}
		\right.
	\end{equation*}
\end{minipage}
\hfill
\begin{minipage}{0.3\textwidth}
	\begin{subequations}
		\label{eqs:m0bar_pert}
		\begin{align}
			&\begin{aligned}
				&\mbox{}\\
				&\mbox{}
			\end{aligned}
			\abs{a} \ll \dm_{21}\\
			&\begin{aligned}
				&\mbox{}\\
				&\mbox{}
			\end{aligned}
			\dm_{21} \ll \abs{a}
		\end{align}
	\end{subequations}
\end{minipage}

\noindent
adequate when looking at the CPT-invariant (vacuum) limit or the T-invariant limit, respectively. 
Notice that both $\mo$ and $\mobar$ converge to the same asymptotic
limit in $\abs{a}$ above $\dm_{21}$.

At the expense of losing precision, 
the evolution of these parameters between the two limits
is illustrated by the approximate interpolations
\begin{equation}
	\label{eq:m0_1st}
	\mo \approx \abs{U_{e1}}^2\, \frac{a\, \dm_{21}}{a+\dm_{21}}\,,
	\hspace{1cm}
	\mobar \approx \dm_{21} - \abs{U_{e2}}^2\, \frac{\abs{a}\, \dm_{21}}{\abs{a}+\dm_{21}}\,,
\end{equation}
with errors $\sim 20\%$, that roughly describes their behavior: 
their vacuum limits are $\mo \to 0$ and $\mobar \to \dm_{21}$,
both vanish when $\dm_{21}$ goes to zero (since $\mo\leq\dm_{21}$)
and go to $\approx \abs{U_{e1}}^2 \dm_{21}$ above $\dm_{21}$.
Notice that first-order approximations in Eqs.~(\ref{eq:m0_1st})
reproduce all four limits in Eqs.~(\ref{eqs:m0_pert},\ref{eqs:m0bar_pert}) 
if one neglects $\abs{U_{e3}}^2$ terms.

The eigenstates in Eq.~(\ref{eq:nutildegeneral}) up to leading order, 
together with the eigenvalues in Eqs.~(\ref{eqs:dmpm_1st}),
reduce to the simple expressions
\begin{equation}
	\ket{\tilde \nu_1} =
	\frac{1}{N_1}
	\mqty[ 
	1\\ 
	\, \frac{\mo -a \abs{U_{e1}}^2}{a U_{e1}^* U_{e2}} \, \\
	0 
	]
	\,,
	\hspace{1cm}
	\ket{\tilde \nu_2} =
	\frac{1}{N_2}
	\mqty[ 
	\, \frac{a U_{e1}^* U_{e2}}{\dm_{21} - \mo + a\abs{U_{e2}}^2} \, \\
	1\\ 
	0 
	]
	\,,
	\hspace{1cm}
	\ket{\tilde \nu_3} =
	\mqty[ 
	0\\ 
	0\\ 
	\, 1  \,
	]
	\,,
\end{equation}
valid for both hierarchies. Antineutrino eigenstates are given by
\begin{equation}
	\ket{\tilde {\bar \nu}_1} =
	\frac{1}{\bar N_1}
	\mqty[ 
	1\\ 
	\, -\frac{\dm_{21}-\mobar -a\abs{U_{e2}}^2}{aU_{e1}U_{e2}^*}\, \\
	0 
	]
	\,,
	\hspace{1cm}
	\ket{\tilde {\bar \nu}_2} =
	\frac{1}{\bar N_2}
	\mqty[ 
	\, -\frac{aU_{e1}U_{e2}^*}{\mobar + a\abs{U_{e1}}^2}\, \\
	1\\ 
	0 
	]
	\,,
	\hspace{1cm}
	\ket{\tilde {\bar \nu}_3} =
	\mqty[ 
	0\\ 
	0\\ 
	\, 1  \,
	]
	\,.
\end{equation}

These eigenstates determine the columns of the $V$ mixing matrix
between matter and vacuum mass eigenstates, 
which allows us to write $\tilde U = UV$ as

\noindent
\begin{minipage}{0.5\textwidth}
	\begin{align*}
		\tilde U_{\alpha 1} &= \frac{1}{N_1}
		\left[ U_{\alpha 1} + \frac{\mo -a \abs{U_{e1}}^2}{a U_{e1}^* U_{e2}}\,
		U_{\alpha 2} \right]\,,\\
		\tilde U_{\alpha 2} &= \frac{1}{N_2}
		\left[ U_{\alpha 2} + \frac{a U_{e1}^* U_{e2}}{\dm_{21} - \mo +
		a\abs{U_{e2}}^2}\, U_{\alpha 1} \right]\,,\\
		\tilde U_{\alpha 3} &= U_{\alpha 3} \,,\\
	\end{align*}
\end{minipage}
\hfill
\begin{minipage}{0.5\textwidth}
	\begin{align}
		\nonumber
		\tilde {\bar U}_{\alpha 1} &= \frac{1}{\bar N_1}
		\left[ U_{\alpha 1}^* -\frac{\dm_{21}-\mobar -a\abs{U_{e2}}^2}{aU_{e1}U_{e2}^*} \,
		U_{\alpha 2}^* \right]\,,\\
		\nonumber
		\tilde {\bar U}_{\alpha 2} &= \frac{1}{\bar N_2}
		\left[ U_{\alpha 2}^* -\frac{aU_{e1}U_{e2}^*}{\mobar + a\abs{U_{e1}}^2}\, 
		U_{\alpha 1}^* \right]\,,\\
		\label{eqs:Utilde}
		\tilde {\bar U}_{\alpha 3} &= U_{\alpha 3}^* \,.\\
		\nonumber
	\end{align}
\end{minipage}

\noindent 
Since matter effects do not depend on all elements of $U_\mathrm{PMNS}$ but
only $U_{ei}$, the above expressions are simpler in the $\alpha = e$ case.
In particular, notice that both
\begin{equation}
	\label{eq:transmutation}
	\tilde U_{e1} = \frac{U_{e1}}{N_1}\,
	\frac{\mo}{a \abs{U_{e1}}^2}
	\,, \hspace{1cm}
	\tilde {\bar U}_{e2} = \frac{U_{e2}^*}{\bar N_2}\,
	\frac{\mobar}{\mobar +a \abs{U_{e2}}^2}
	\,,
\end{equation}
vanish if $\dm_{21}=0$ for all $a$.
This fact originates in the transmutation~\cite{matter0th}
of masses in vacuum to mixings in matter, leading to 
the absence of genuine CP violation in matter if $\dm_{21}=0$,
even though there are three non-degenerate neutrino masses.

These expressions reproduce the right vacuum limit, as seen by developing
$\abs{a}\ll\dm_{21}$,
\begin{equation}
	\tilde U_{\alpha 1} = U_{\alpha 1} -
	\frac{a}{\dm_{21}}\, U_{e1}U_{e2}^*U_{\alpha 2} \,,
	\hspace{1cm}
	\tilde U_{\alpha 2} = U_{\alpha 2} + 
	\frac{a}{\dm_{21}}\, U_{e2} U_{e1}^* U_{\alpha 1} \,.
\end{equation}
A surprising result, however, appears when assuming $\dm_{21}\ll \abs{a}$,
\begin{subequations}
	\label{eqs:mixings_higha}
	\begin{align}
		\tilde U_{\alpha 1} &= \frac{\abs{U_{e2}}}{\sqrt{1-\abs{U_{e3}}^2}}
		\left[ U_{\alpha 1} -\frac{U_{e1}}{U_{e2}}\,U_{\alpha 2} 
		+\frac{\dm_{21}\, U_{e1}}{a \left( 1-\abs{U_{e3}}^2\right)^2}
		\left( U_{e1}^* U_{\alpha 1} + U_{e2}^* U_{\alpha 2} \right)\right] \,,\\
		\tilde U_{\alpha 2} &= \frac{\abs{U_{e2}}}{\sqrt{1-\abs{U_{e3}}^2}}
		\left[ U_{\alpha 2} +\frac{U_{e1}^*}{U_{e2}^*}\,U_{\alpha 1} 
		-\frac{\dm_{21}\, U_{e1}^*}{a \left( 1-\abs{U_{e3}}^2\right)^2}
		\left( U_{e2} U_{\alpha 1} - U_{e1} U_{\alpha 2} \right)\right] \,,
	\end{align}
\end{subequations}
showing that 
$\lim\limits_{a \to 0} \lim\limits_{\dm_{21} \to 0} \tilde U
\neq U_\mathrm{PMNS} =
\lim\limits_{\dm_{21} \to 0} \lim\limits_{a \to 0} \tilde U$
! This is strongly illustrated in the case 
$\lim\limits_{\dm_{21} \to 0} \tilde U_{e1} = 
\lim\limits_{\dm_{21} \to 0} \tilde {\bar U}_{e2} = 0\, \forall a$.
The reason behind this subtlety is the following.
Setting $\dm_{21}=0$ in vacuum means that $\nu_1$ and $\nu_2$ are degenerate.
Therefore, any two independent linear combinations of them can be chosen as basis states, 
which
in the language of the standard parametrization would mean that $\theta_{12}$
is nonphysical. Adding the matter potential to this system breaks the
degeneracy: the arbitrariness in $\theta_{12}$ is lost in favor of the
eigenstates of the perturbation. Since the matter term in the neutrino Hamiltonian adds
$a>0$ to the $e$-flavor component, this fact results in $\tilde \nu_1$ and $\tilde
\nu_2$ such that $\tilde \nu_2$ is mainly $\nu_e$, forcing the $\tilde U_{e1} = 0$ we
obtained.
The change of sign in $a$ for the antineutrino case forces analogously 
$\tilde {\bar \nu}_1$ to be mainly $\bar \nu_e$, explaining the limit
$\tilde {\bar U}_{e2} =0$.

\begin{figure}[t]
	\centering
	\begin{tikzpicture}[line width=1 pt, scale=1.5]
		\node at (0,0){
			\includegraphics[width=0.475\textwidth]{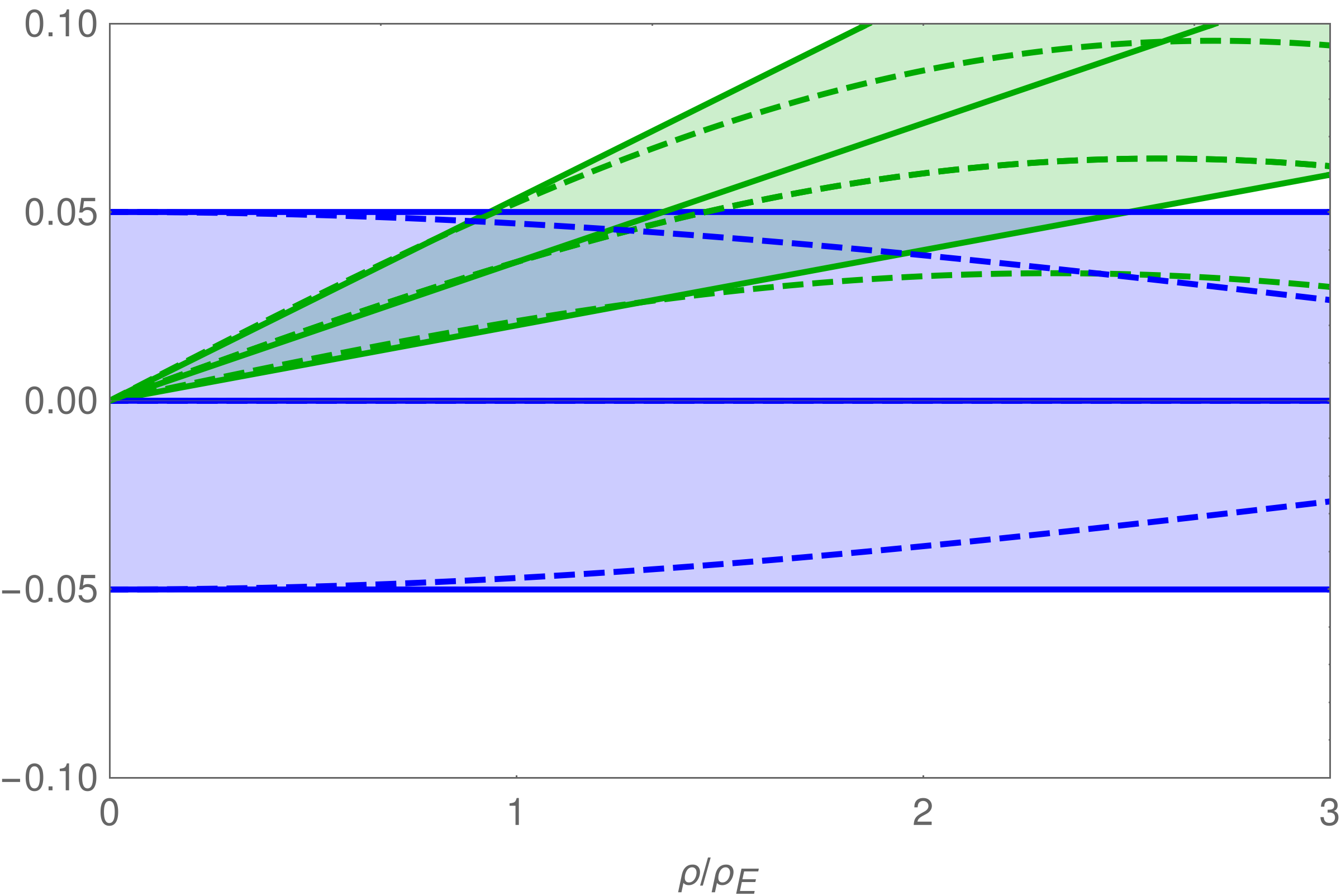}
		};

		\node at (-2.0,1.5)[below right]{NH};
	\end{tikzpicture}
	\hfill
	\begin{tikzpicture}[line width=1 pt, scale=1.5]
		\node at (0,0){
			\includegraphics[width=0.475\textwidth]{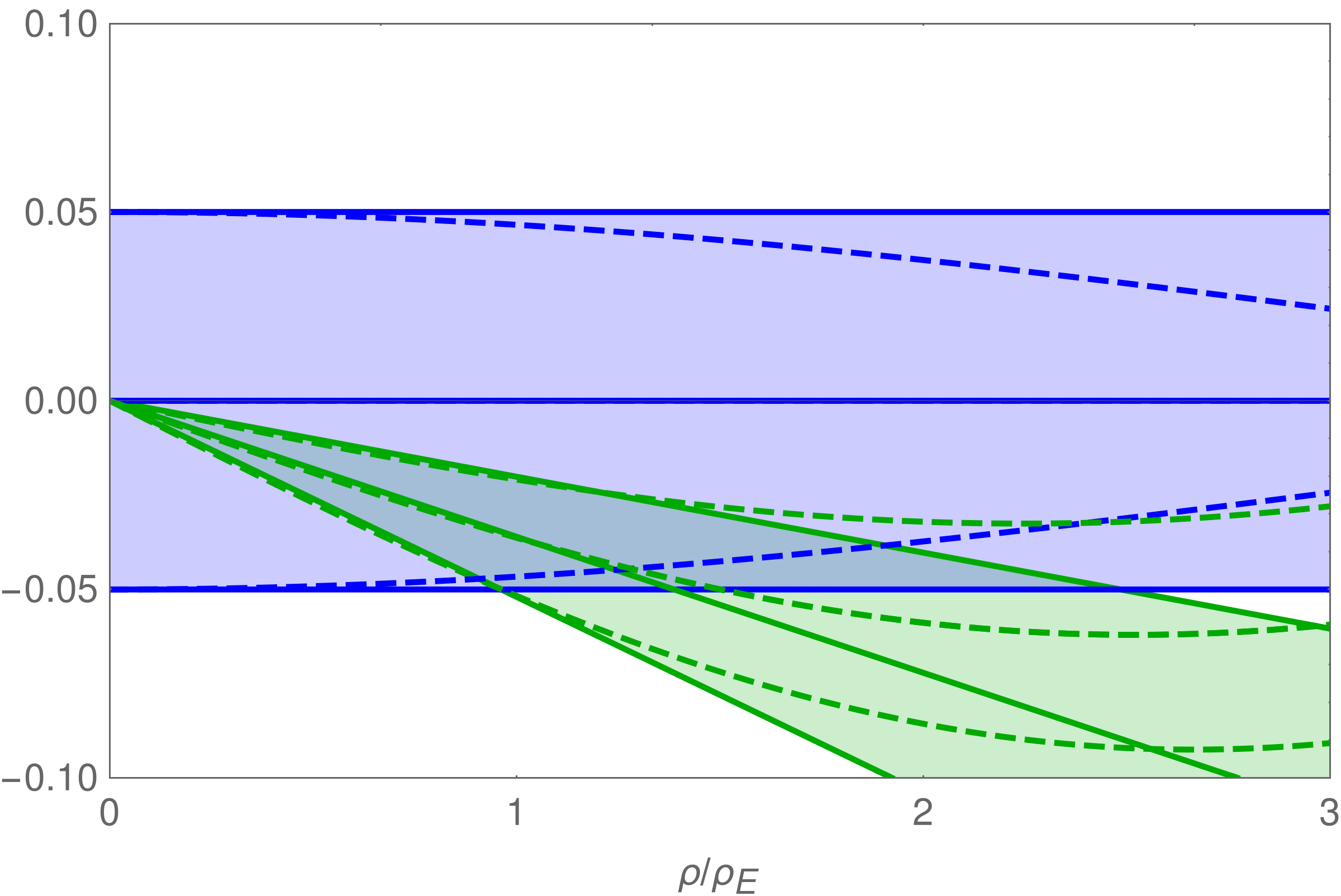}
		};

		\node at (-2.0,1.5)[below right]{IH};
	\end{tikzpicture}
	\hfill
	\caption{CPT-odd (green) and T-odd (blue) components of $\asym{CP}_{\mu e}$
	as functions of the matter density $\rho$ in units of that of the Earth mantle,
	at fixed $(E,\, L)=(0.75~\mathrm{GeV},\, 1300~\mathrm{km})$.
	Both the exact (dashed) and the analytical (solid)
	results from Eqs.(\ref{eqs:vacuum_masses}, \ref{eqs:vacuum_mixings}) are shown.
	Normal Hierarchy in the left pannel, Inverted Hierarchy in the right pannel.
	The bands correspond to all possible values changing $\delta$ in $(0,\,2\pi)$;
	%the lines are $\delta = 0^\circ,\, 90^\circ,\, 180^\circ,\, 270^\circ$.
	the upper/central/lower lines for $\asym{CPT}_{\mu e} (\asym{T}_{\mu e})$ 
	correspond to $\cos\delta (\sin\delta) = -1,\, 0,\, 1$.
	}
	\label{fig:Asyms_vacuum}
\end{figure}

This behavior shows that the vacuum connection should be analyzed in the
regime where $\abs{a} \ll \dm_{21} \ll \abs{\dm_{31}}$.
The definite $a$-parity of the two components of the CP asymmetry defined in
the previous Section forces the leading-order term in $\asym{T}_{\alpha\beta}$
to be independent of $a$, whereas $\asym{CPT}_{\alpha\beta}$ is linear.
To provide a precise description of $\asym{CPT}_{\alpha\beta}$ in this region,
we keep $\abs{U_{e3}}^2$ terms in the leading order, as well as
all linear terms in $a/\dm_{21}$ and $a/\dm_{31}$ 
in both the mass squared differences,
\begin{equation}
	\label{eqs:vacuum_masses}
	\dmt_{21} \approx \dm_{21} - a(\abs{U_{e1}}^2-\abs{U_{e2}}^2) \,,
	\hspace{0.6cm}
	\dmt_{31} \approx \dm_{31} - a\abs{U_{e1}}^2\,,
	\hspace{0.6cm}
	\dmt_{32} \approx \dm_{32} - a\abs{U_{e2}}^2\,,\\
\end{equation}
%valid also for antineutrinos changing $a\to -a$,
and the mixings,
\begin{subequations}
	\label{eqs:vacuum_mixings}
	\begin{align}
		\tilde U_{\alpha 1} &= U_{\alpha 1}
		-\frac{a\,U_{e1}}{\dm_{21}}\, U_{e2}^*U_{\alpha 2}
		-\frac{a\,U_{e1}}{\dm_{31}}\, U_{e3}^*U_{\alpha 3} \,,\\
		\tilde U_{\alpha 2} &= U_{\alpha 2}  
		+\frac{a\,U_{e2}}{\dm_{21}}\, U_{e1}^* U_{\alpha 1}
		-\frac{a\,U_{e2}}{\dm_{31}}\, U_{e3}^* U_{\alpha 3} \,,\\
		\tilde U_{\alpha 3} &= U_{\alpha 3}  
		+\frac{a\,U_{e3}}{\dm_{31}}\, U_{e1}^* U_{\alpha 1}
		+\frac{a\,U_{e3}}{\dm_{31}}\, U_{e2}^* U_{\alpha 2} \,.
	\end{align}
\end{subequations}
The CP asymmetry components $\asym{CPT}_{\mu e}$ and $\asym{T}_{\mu e}$ computed using these
expressions are represented in Fig.~\ref{fig:Asyms_vacuum},
compared with the exact results, for both hierarchies
at fixed $E$ and $L$ as functions of the matter potential.

The analytic approximations for constant $\asym{T}_{\mu e}$ and linear
$\asym{CPT}_{\mu e}$ work well at low matter densities, as they should, but
their range of validity is much larger than expected. 
For the values used in the Figure, the point $a = \dm_{21}$ corresponds to
$\rho = 0.44\rho_E$, so the previous expansions should only work for
$\rho \ll 0.44\rho_E$. The fact that they work reasonably well even above
$\rho_E$ hints that higher-order corrections are dominated by $(a/\dm_{31})^2$.

This surprising feature stems from the fact 
that corrections $(a/\dm_{21})^2$
are inoperative 
in the region $\abs{a} \ll \dm_{21} \ll \abs{\dm_{31}}$
for the 
$\asym{CPT}_{\mu e}$ and $\asym{T}_{\mu e}$ observables.
This behavior is explained by peculiar dependence on the mixings and masses of
the oscillation probabilities, as can be understood from the matter-vacuum invariants we will
exploit in the following Sections for both
Im$\tilde J^{ij}_{\alpha\beta}$ (Section~\ref{sec:AT})
and
Re$\tilde J^{ij}_{\alpha\beta}$ (Section~\ref{sec:Hier}).
As will be discussed, they lead to dependencies in the oscillation
probabilities in Eq.~(\ref{eq:Pab}) on the phases associated to the small quatities $a$ and
$\dm_{21}$ of the form $\frac{1}{\Delta}\sin \Delta$, which cancel out if both
of them are small, independently of whether $\abs{a}\ll \dm_{21}$ or
$\dm_{21}\ll \abs{a}$.
This cancellation will happen as long as $\Delta = \frac{\epsilon L}{4E} \ll
1$,
for $\epsilon = a,\, \dm_{21}$.
This peculiar dependence in the oscillation probabilities is responsible for
the restoration of the commutability of the limits $a\to 0$ and $\dm_{21}\to 0$
at this level, even though they do not commute at the mixings level.

\subsection{Actual experiments: fixed $L$ in the Earth mantle and variable $E$}
\label{sec:actual_experiments}

In the previous Subsection we discussed the way to obtain analytic approximated
expressions for neutrino oscillations in matter that reproduce the right vacuum
limit, i.e. the limit when the matter parameter $a\to 0$ at fixed energy due to
the matter density going to zero.

In the following we consider the constant value of the matter density in the
Earth mantle~\cite{a=3E}, and discuss dependencies in $a$ as dependencies in the
neutrino energy in $\nu_\mu \to \nu_e$ transitions. 
In fact, the actual best-fit value~\cite{mariam} for
$\dm_{21}$ shows that the relation between $a$ and $\dm_{21}$ is given by
$\abs{a} \approx 3 (E/\mathrm{GeV})\dm_{21}$, 
so we can use Eqs.~(\ref{eqs:Utilde}) from the previous Subsection expanding up
to second order in $\dm_{21}/a$, with errors only $\sim 3\%$ around 1 GeV.

As in Eq.~(\ref{eq:Pab}), all observable quantities can be written in terms of
the rephasing-invariant mixings $\tilde J_{\alpha\beta}^{ij}$. Since $\tilde
U_{e1}$ in Eq.~(\ref{eq:transmutation}) is a first order quantity, 
as we discussed, and expanding up to second order also in
$\abs{U_{e3}} \ll 1$, 
which is of the same size than $\dm_{21}/a$, 
we find that all $\tilde J_{e \alpha}^{ij}$ can be
calculated at second order in these two quantities
using our first-order $\tilde U_{\alpha i}$
in Eqs.~(\ref{eqs:mixings_higha}),
\begin{subequations}
	\begin{align}
		\tilde J_{e \alpha}^{13} &= 
		\frac{\dm_{21}}{a} \left( \abs{U_{e2}}^2 J_{e\alpha}^{13} - 
		\abs{U_{e1}}^2 J_{e\alpha}^{23}\right) 
		%+ \frac{a}{\dm_{31}}\, J_{e\alpha}^{13} 
		\,,\\
		\tilde J_{e \alpha}^{23} &= J_{e\alpha}^{23} + J_{e\alpha}^{13}
		-\frac{\dm_{21}}{a} \left( \abs{U_{e2}}^2 J_{e\alpha}^{13} - 
		\abs{U_{e1}}^2 J_{e\alpha}^{23}\right) 
		%+ \frac{a}{\dm_{31}}\, J_{e\alpha}^{23} 
		\,,\\
		\nonumber
		\tilde J_{e \alpha}^{12} &= \frac{\dm_{21}}{a} \left[ \abs{U_{e2}}^2
		J_{e\alpha}^{12} - \abs{U_{e1}}^2 J_{e\alpha}^{21} + \abs{U_{e1}}^2
		\abs{U_{e2}}^2 \left( \abs{U_{\alpha 1}}^2 - \abs{U_{\alpha 2}}^2 \right) \right]\\  
		&- \left[\frac{\dm_{21}}{a}\right]^2 \abs{U_{e1}}^2\abs{U_{e2}}^2(1-\abs{U_{\alpha 3}}^2)
		\,.
	\end{align}
\end{subequations}

However, as discussed in the previous Subsection, the definite odd $a$-parity of
the CPT component of the CP asymmetry implies that linear terms in
$a/\dm_{31}$ are relevant to describe $\asym{CPT}_{\alpha\beta}$, so we must keep
them as well.
These linear terms can be easily calculated setting $\dm_{21}\to~\!0$  
in the eigenstates in Eq.~(\ref{eq:nutildegeneral}). 
Analogously, we obtain linear corrections
in $\dm_{21}/\dm_{31}$ to the previous $\tilde J_{\alpha\beta}^{ij}$ setting $a\to 0$
in the eigenstates.
The resulting rephasing-invariant mixings, written in the standard parametrization for $\alpha = \mu$,
which is the relevant transition for accelerator experiments,
are
\begin{subequations}
	\label{eqs:Jtmue}
	\begin{align}
		\tilde J_{e \mu}^{13} &= 
		-\left(\frac{\dm_{21}}{a} + \frac{\dm_{21}}{\dm_{31}} \right)\, 
		c_{12} c_{13}^2 c_{23} s_{12} s_{13} s_{23}\, e^{i\delta} 
		\,,\\
		\tilde J_{e \mu}^{23} &= 
		-c_{13}^2 s_{13}^2 s_{23}^2 \left( 1+\frac{2a}{\dm_{31}}\right)
		+ \left(\frac{\dm_{21}}{a} + \frac{\dm_{21}}{\dm_{31}} \right)\,
		c_{12} c_{13}^2 c_{23} s_{12} s_{13} s_{23}\, e^{i\delta} 
		\,,\\
		\tilde J_{e \mu}^{12} &= 
		\left( \frac{\dm_{21}}{a} + \frac{\dm_{21}}{\dm_{31}} \right)\,
		c_{12} c_{13}^2 c_{23} s_{12} s_{13} s_{23}\, e^{i\delta}
		-\left[\frac{\dm_{21}}{a} \right]^2 c_{12}^2 s_{12}^2 c_{23}^2 
		\,.
	\end{align}
\end{subequations}
We find in the rephasing-invariant mixings of Eqs.~(\ref{eqs:Jtmue}) the four
observable reparametrization invariants 
$\mathcal{J} \equiv J_r \sin\delta = c_{12} c_{13}^2 c_{23} s_{12} s_{13} s_{23} \sin\delta$,
$R \equiv J_r \cos\delta $, %= c_{12} c_{13}^2 c_{23} s_{12} s_{13} s_{23} \cos\delta$,
$S \equiv c_{13}^2 s_{13}^2 s_{23}^2$ and
$T \equiv c_{12}^2 s_{12}^2 c_{23}^2$.
Notice that these are the same results found in Ref.~\cite{cervera} after
further expanding in $\abs{a}\ll\abs{\dm_{31}}$, as expected.
Since $\abs{\dm_{31}} \approx 33 \dm_{21}$, it turns out that 
$\abs{\dm_{31}} \approx 11 \abs{a}/ (E/\mathrm{GeV})$, 
so expanding in $\abs{a}\ll\abs{\dm_{31}}$ around the $E\sim$ GeV
region is as reasonable as expanding in $\abs{U_{e3}} \ll 1$.
All $\tilde J_{\mu e}^{ij}$ are already second order
in $\dm_{21}$ and $\abs{U_{e3}}$, so
we can neglect them    
in the oscillation arguments,
\begin{alignat}{3}
	\nonumber
	&\dmt_{21} \approx a\,,
	\hspace{1cm}
	&&\dmt_{31} \approx \dm_{31}\,,
	\hspace{1cm}
	&&\dmt_{32} \approx \dm_{31}-a \,,\\
	\label{eq:masses_exp}
	&\dmtbar_{21} \approx \abs{a}\,,
	\hspace{1cm}
	&&\dmtbar_{31} \approx \dm_{31} + \abs{a}\,,
	\hspace{1cm}
	&&\dmtbar_{32} \approx \dm_{31} \,.
\end{alignat}
In this regime, the only oscillation phases are
the vacuum phase $\Delta_{31} \propto L/E$ and the %The phase $\tilde \Delta_{21}$ has the
constant (for a given baseline through the Earth mantle) %value
\begin{equation}
	A\equiv \frac{aL}{4E} = 3.8\, \dm_{21}(\mathrm{eV}^2)\, L(\mathrm{km})
	= 0.29\, \frac{L}{1000~\mathrm{km}}\,.
\end{equation}
This value is not particularly small at long baselines, but we remind the
reader that both $\asym{CPT}_{\mu e}$ and $\asym{T}_{\mu e}$ have definite
parity in $a$, the first one being odd and the second one even, 
as we proved in Section~\ref{sec:theorem}. This means that
corrections to the leading order in each component of the CP asymmetry will be
quadratic in $a$, and so we can also expand up to leading order.

In summary, the expansion quantities used 
are the phase $A$ and,
up to second order,
\begin{equation}
	\frac{\dm_{21}}{\dm_{31}} \sim 0.030\,,
	\hspace{0.8cm}
	\frac{\dm_{21}}{a} \sim \frac{0.34}{E/\mathrm{GeV}}\,,
	\hspace{0.8cm}
	\frac{a}{\dm_{31}} \sim 0.091\, E/\mathrm{GeV}\,,
	\hspace{0.8cm}
	\abs{U_{e3}} \sim 0.15\,.
\end{equation}
%up to second order in $\dm_{21}/a$ or equivalent accuracy.

Taking into account the rephasing-invariant mixings~(\ref{eqs:Jtmue}),
with the symmetry property $\tilde J_{\mu e}^{ij} = \tilde J_{e\mu}^{ji}$,
and the mass differences in matter~(\ref{eq:masses_exp}), we find

\noindent
\boxed{
	\begin{minipage}{0.99\textwidth}
		\vspace{-0.4cm}
		\begin{subequations}
			\label{eqs:analAsyms}
			\begin{align}
				\label{eq:analACPT}
				\asym{CPT}_{\mu e} &=
					16\, A 
					\left[ \frac{\sin\Delta_{31}}{\Delta_{31}}-\cos\Delta_{31} \right]
					\left( S\sin\Delta_{31} + J_r\cos\delta\, \Delta_{21}\cos\Delta_{31} \right)
					+\mathcal{O}(A^3)\,,\\
				\label{eq:analAT}
				\asym{T}_{\mu e} &=
					-16\,J_r \sin\delta\, \Delta_{21}\sin^2\Delta_{31}
					+\mathcal{O}(A^2) \,,
			\end{align}
		\end{subequations}
		\vspace{-0.7cm}
	\end{minipage}
}

\noindent
where $S \equiv c_{13}^2 s_{13}^2 s_{23}^2$,
$J_r \equiv c_{12} c_{13}^2 c_{23} s_{12} s_{13} s_{23}$,
$A\equiv \frac{aL}{4E} \propto L$ and
the two $\Delta_{ij} \equiv \frac{\dm_{ij} L}{4E} \propto~\!L/E$.
From these expressions, which are precise enough to provide understanding of the
physics behind these observables, we find that $\asym{T}_{\mu e}$ in matter is
well described by its vacuum value. Since $\Delta_{21}$ is small, this means
that $\asym{T}_{\mu e}$ oscillates as $\frac{1}{E}\, \sin^2\Delta_{31}$.
$\asym{CPT}_{\mu e}$, which vanishes when
$a\to 0$, is very well described by its leading (first) order in $a$.

\begin{figure}[b!]
	\centering
	\begin{tikzpicture}[line width=1 pt, scale=1.5]
		\node at (0,0){
			\includegraphics[width=0.475\textwidth]{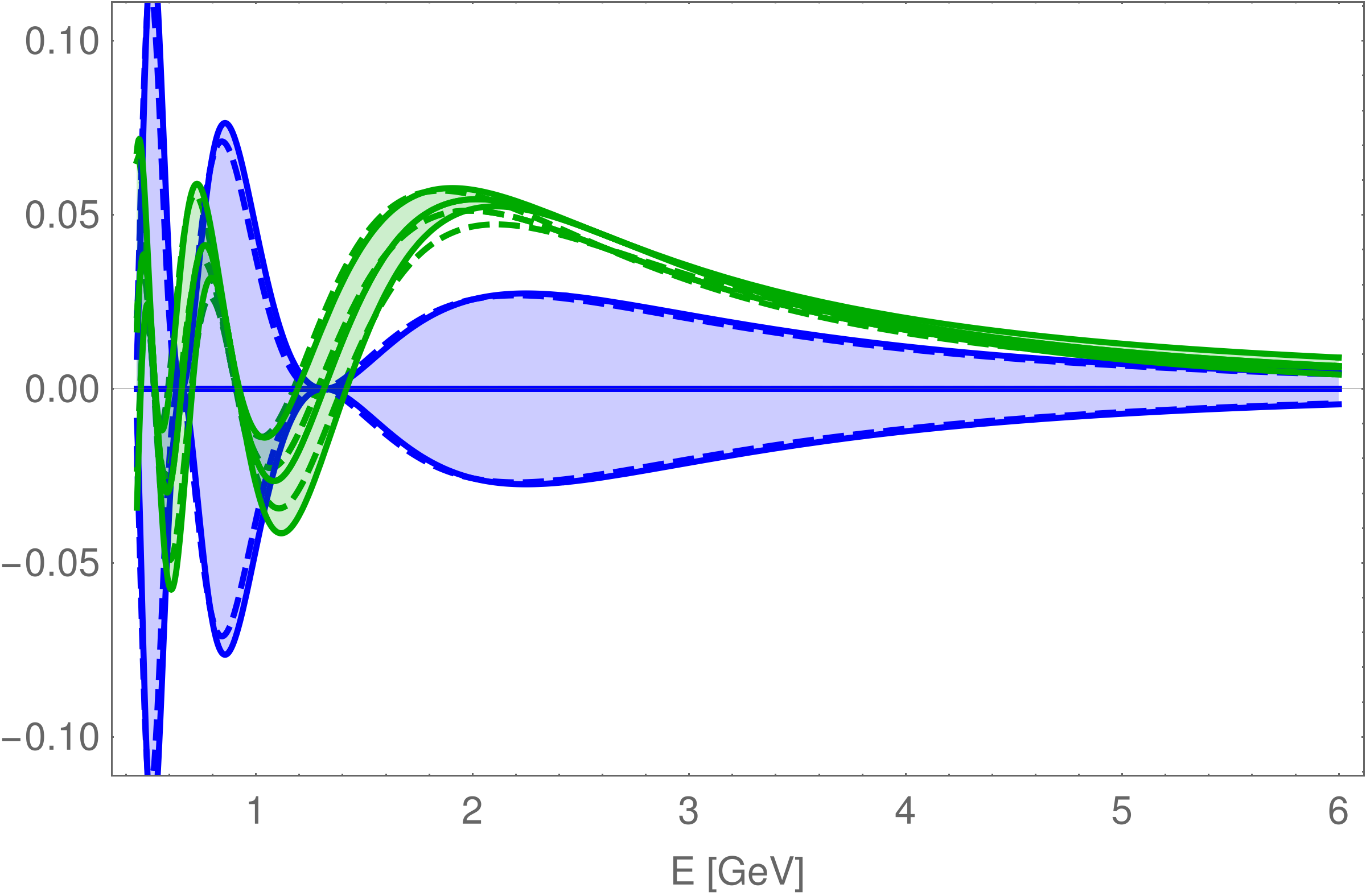}
		};

		\node at (2.4,1.50)[below left]{NH};
	\end{tikzpicture}
	\hfill
	\begin{tikzpicture}[line width=1 pt, scale=1.5]
		\node at (0,0){
			\includegraphics[width=0.475\textwidth]{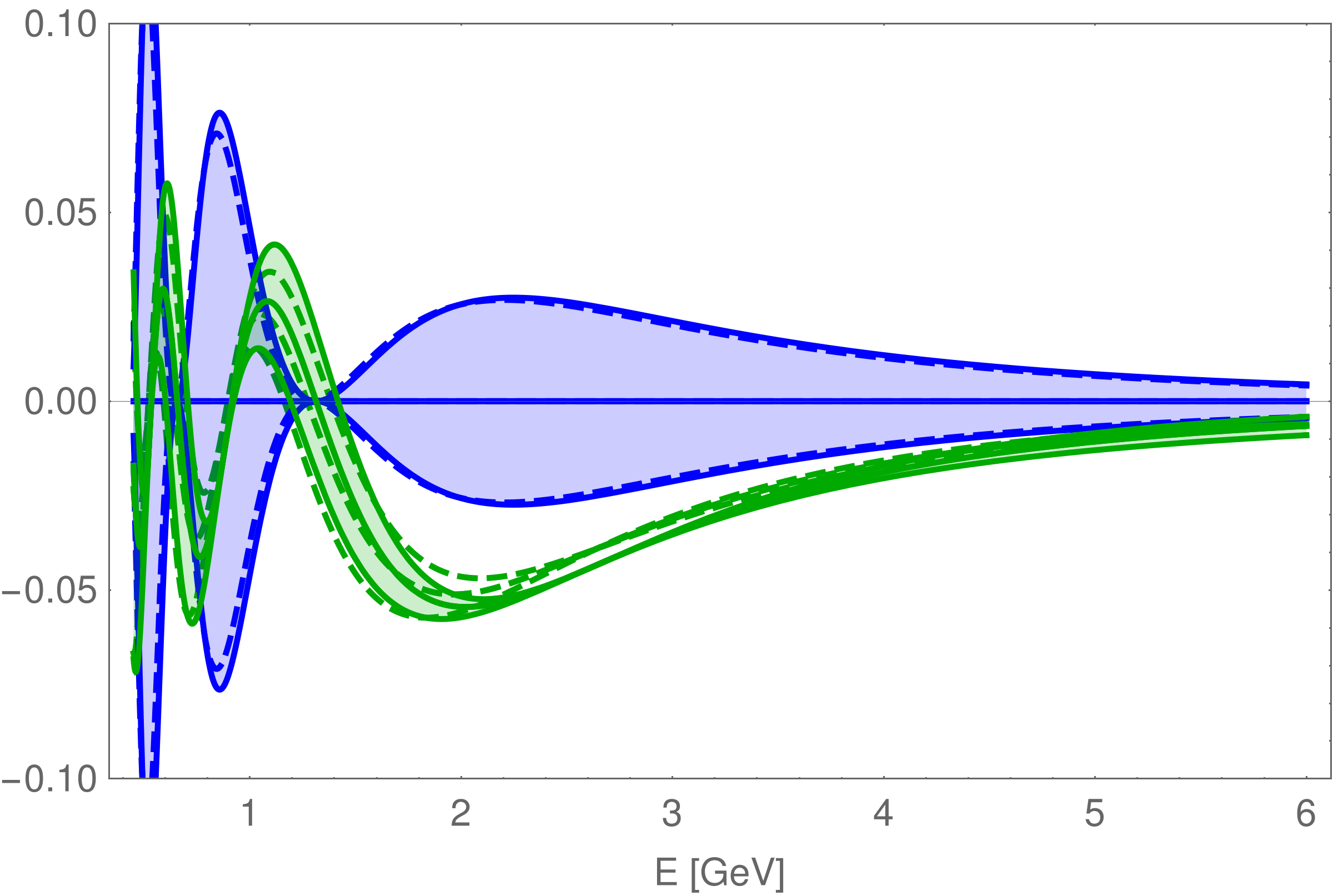}
		};

		\node at (2.4,1.50)[below left]{IH};
	\end{tikzpicture}
	\caption{CPT-odd (green) and T-odd (blue) components of $\asym{CP}_{\mu e}$
	as functions of the neutrino energy $E$ through the Earth mantle (of constant density)
	and a baseline of $L = 1300$~km.
	Both the exact (dashed) and the analytical (solid)
	results from Eqs.(\ref{eqs:analAsyms}) are shown.
	%Normal Hierarchy in the left pannel, Inverted Hierarchy in the right pannel.
	Normal/Inverted Hierarchy in the left/right pannel.
	The bands correspond to all possible values changing $\delta$ in $(0,\,2\pi)$;
	the upper/central/lower lines for $\asym{CPT}_{\mu e} (\asym{T}_{\mu e})$ 
	correspond to $\cos\delta (\sin\delta) = -1,\, 0,\, 1$.
	}
	\label{fig:Asyms}
\end{figure}

The agreement of Eqs.~(\ref{eqs:analAsyms}) with the exact result is shown in
Fig.~\ref{fig:Asyms}, which makes clear that, even if the value of the
asymmetries in the maxima are a bit off, their position and the general
behavior are well reproduced.
Therefore, Eqs.~(\ref{eqs:analAsyms}) are the perfect tool to understand 
the energy dependencies of the two disentangled components 
$\asym{CPT}_{\mu e}$ and $\asym{T}_{\mu e}$
of the CP asymmetry and search for their actual experimental separation.

\section{A closer look at the genuine CPV component}
\label{sec:AT}

The last term in Eq.~(\ref{eq:Hf}) indicates that the Hamiltonian of
our problem in the flavor basis is proportional to the hermitian
mass matrix squared in matter
\begin{equation}
	2E\, H \equiv H' = \tilde M_\nu \tilde M_\nu^\dagger \,.
\end{equation}
In such a basis, the necessary and sufficient condition for CP
invariance is~\cite{CP-condition}
\begin{equation}
	\label{eq:CP-condition}
	\mathrm{Im}[H'_{e\mu} H'_{\mu\tau} H'_{\tau e} ] = 0\,.
\end{equation}
For any flavor-diagonal interaction of neutrinos with matter, 
the last condition is equal to that for neutrino mass matrices in vacuum. 
This invariance~\cite{Harrison-Scott} of the left-hand side of the last equation
(\ref{eq:CP-condition}) between
the CP behavior of neutrinos in vacuum and in matter has far-reaching
consequences for the observable rephasing-invariant mixings of neutrinos 
$\tilde J_{\alpha\beta}^{ij}$ and antineutrinos 
$\tilde {\bar J}_{\alpha\beta}^{ij}$ in matter.

The explicit calculation of the matter-vacuum invariant genuine CP violation expression for neutrinos, 
antineutrinos and in vacuum leads to
\begin{equation}
	\label{eq:invariantImJ}
	\dmt_{12} \dmt_{23} \dmt_{31} \mathcal{\tilde J}
	= \dmtbar_{12} \dmtbar_{23} \dmtbar_{31} \mathcal{\tilde {\bar J}}
	= \dm_{12} \dm_{23} \dm_{31} \mathcal{J}\,,
\end{equation}
where $\mathcal{J}$ is the rephasing-invariant CPV quantity in
vacuum~\cite{Jarlskog},
$\mathcal{J}\! =\! c_{12} c_{13}^2 c_{23} s_{12} s_{13} s_{23} \sin\delta$.
The proportionality of $\mathcal{\tilde J}$ and $\mathcal{\tilde {\bar J}}$ 
to $\dm_{21}$
explains the absence of genuine CP violation in matter in the
limit of vanishing $\dm_{21}$, even in the presence of three
non-degenerate neutrinos and antineutrinos in matter. The vanishing 
of $\mathcal{\tilde J}$ and $\mathcal{\tilde {\bar J}}$ in this limit comes
from
the transmutation of masses in vacuum to mixings in matter 
calculated in Section~\ref{sec:vacuum_limit}, leading to $\tilde U_{e1}=0$
and $\tilde {\bar U}_{e2}=0$. To leading order in $\dm_{21}$, the non-vanishing
$\mathcal{\tilde J}$ and $\mathcal{\tilde {\bar J}}$ differ by linear terms in the 
matter potential $a$ present in the neutrino masses in matter. 

\begin{figure}[t]
	\centering
	\begin{tikzpicture}[line width=1 pt, scale=1.5]
		\node at (0,0){
			\includegraphics[width=0.475\textwidth]{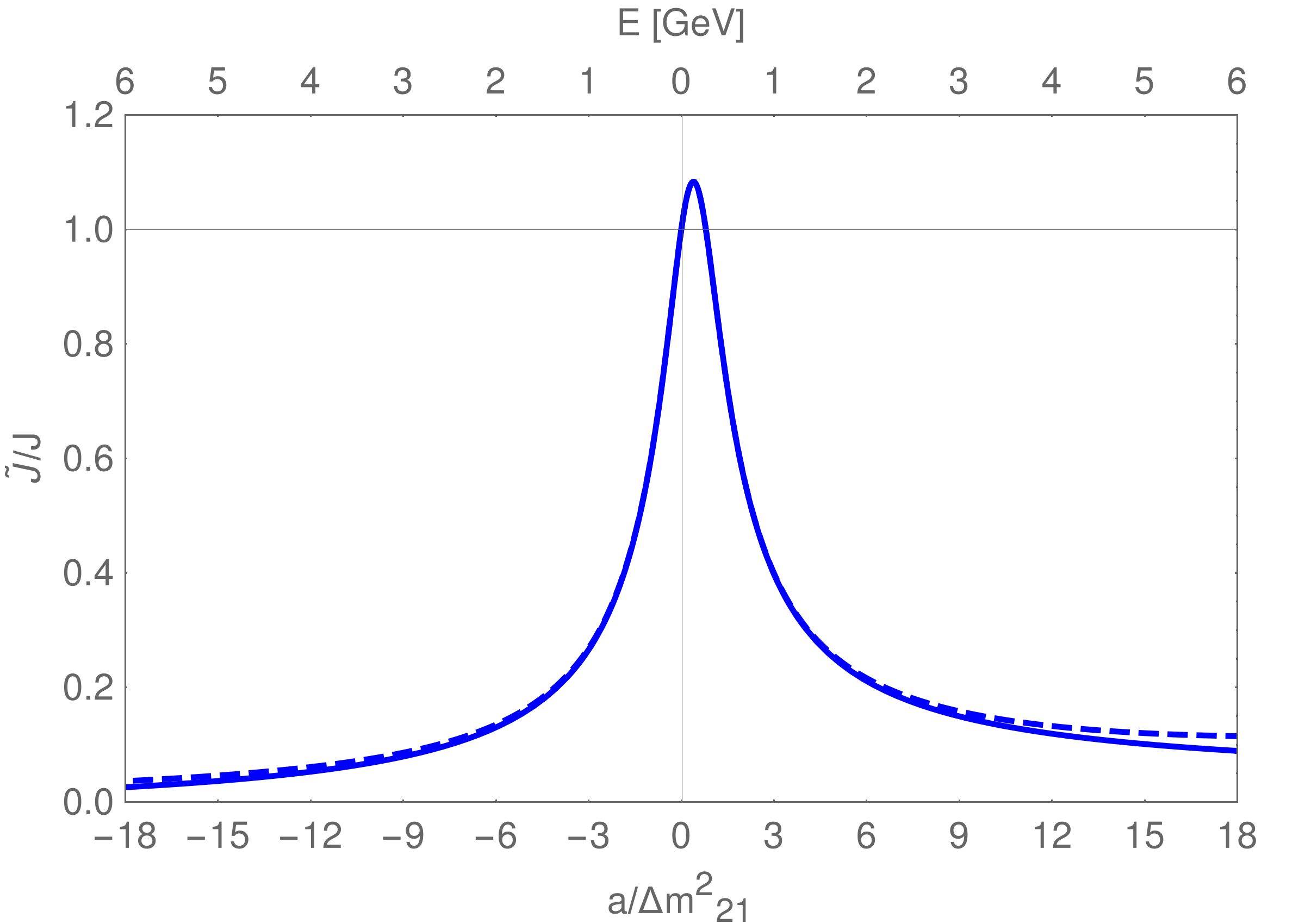}
		};

		\node at (2.15,1.25)[below left]{NH};
	\end{tikzpicture}
	\hfill
	\begin{tikzpicture}[line width=1 pt, scale=1.5]
		\node at (0,0){
			\includegraphics[width=0.475\textwidth]{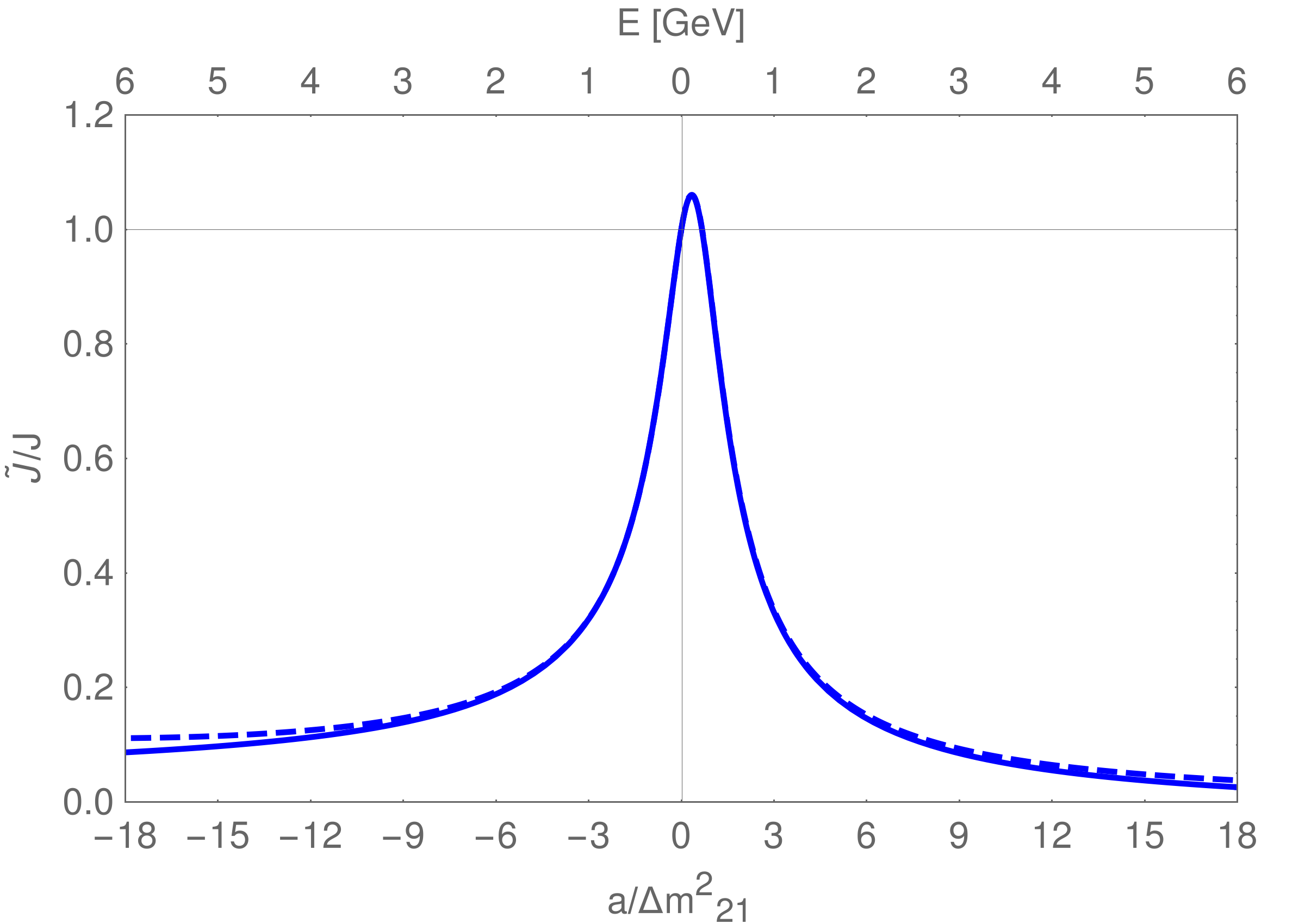}
		};

		\node at (2.15,1.25)[below left]{IH};
	\end{tikzpicture}
	\caption{
		$\mathcal{\tilde J}/\mathcal{J}$ ratio
	for both neutrinos $(a>0)$ and antineutrinos $(a < 0)$. 
	The horizontal axis shows both the evolution of the matter parameter $a$ at
	fixed energy (lower labels), i.e. changing the matter density,
	and as function of the energy if the constant
	density is chosen as that of the Earth mantle (upper labels).
	Both the exact (dashed) and the analytical (solid)
	results from Eqs.(\ref{eqs:analImJ}) are shown.
	Normal Hierarchy in the left pannel,
	Inverted Hierarchy in the right pannel.}
	\label{fig:exp_ImJratio}
\end{figure}

Using the analytic perturbation expansion of Section~\ref{sec:desarrollos} for the
connection between quantities in matter and in vacuum, we can write
\begin{subequations}
	\label{eqs:analImJ}
	\begin{align}
		\mathcal{\tilde J} &=
			\frac{\dm_{21} \left[ \dm_{31} +a%(1-3\abs{U_{e3}}^2) 
			\right]}%
			{\dm_{31} \left[ \dm_{21} - 2\mo +a(1-\abs{U_{e3}}^2) \right]}\, 
			\mathcal{J} \,,
			\hspace{1cm}
			a\geq 0\,,\\
		\mathcal{\tilde {\bar J}} &=
			\frac{\dm_{21} \left[ \dm_{31} +a%(1-\abs{U_{e3}^2}) 
			\right]}%
			{\dm_{31} \left[2\mobar -\dm_{21} -a(1-\abs{U_{e3}}^2) \right]}\, 
			\mathcal{J} \,,
			\hspace{1cm}
			a\leq 0\,,
	\end{align}
\end{subequations}
Notice that the proportionality factors in Eqs.~(\ref{eqs:analImJ}) are neutrino
energy dependent through $a$,
as shown in Fig.~\ref{fig:exp_ImJratio}. 
The behavior at low/high energies can be easily understood using the expansions
at leading order of $\mo$ and $\mobar$ in
Eqs.~(\ref{eqs:m0_pert},\ref{eqs:m0bar_pert}). 
Indeed, at low energies
\begin{subequations}
	\label{eqs:analImJ_lowE}
	\begin{align}
		\mathcal{\tilde J} &\approx
		\mathcal{J} \left[
			1 + \frac{a (\abs{U_{e1}}^2-\abs{U_{e2}}^2)}{\dm_{21}}
			+\frac{a}{\dm_{31}}
		\right]
		> \mathcal{J} \,,\\
		\mathcal{\tilde {\bar J}} &\approx
		\mathcal{J} \left[
			1 - \frac{\abs{a} (1-\abs{U_{e2}}^2)}{\dm_{21}}
			-\frac{\abs{a}}{\dm_{31}}
		\right]
		< \mathcal{J} \,,
	\end{align}
\end{subequations}
the ratio increases (decreases) with respect to 1 for (anti)neutrinos
independently of sign$(\dm_{31})$ due to $\dm_{21}\ll\abs{\dm_{31}}$,
whereas at high energies
\begin{subequations}
	\label{eqs:analImJ_highE}
	\begin{align}
		\mathcal{\tilde J} &\approx
			\mathcal{J}\;
			\frac{\dm_{21} (\dm_{31}+a)}{a\,\dm_{31}}\,,\\
		\mathcal{\tilde {\bar J}} &\approx
			\mathcal{J}\;
			\frac{\dm_{21} (\dm_{31}-\abs{a})}{\abs{a}\,\dm_{31}}\,,
	\end{align}
\end{subequations}
both of them decrease roughly as $1/a$, and changing the sign$(\dm_{31})$ is equivalent to
changing the sign$(a)$, 
which explains why the two plots in Fig.~\ref{fig:exp_ImJratio} seem to be
symmetrical.

\begin{figure}[t]
	\centering
	\begin{tikzpicture}[line width=1 pt, scale=1.5]
		\node at (0,0){
			\includegraphics[width=0.475\textwidth]{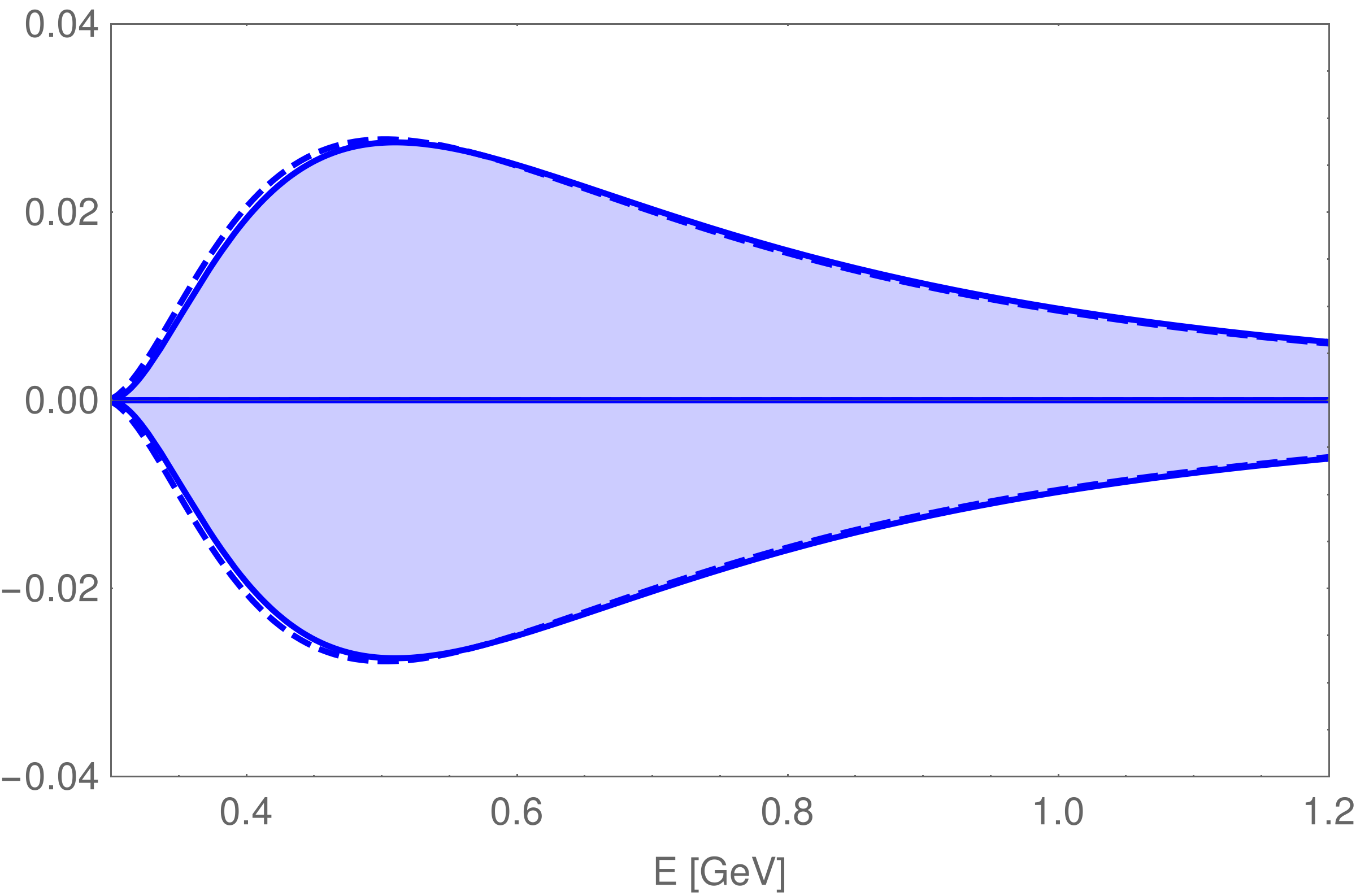}
		};

		\node at (2.35,1.5)[below left]{T2HK};
	\end{tikzpicture}
	\hfill
	\begin{tikzpicture}[line width=1 pt, scale=1.5]
		\node at (0,0){
			\includegraphics[width=0.475\textwidth]{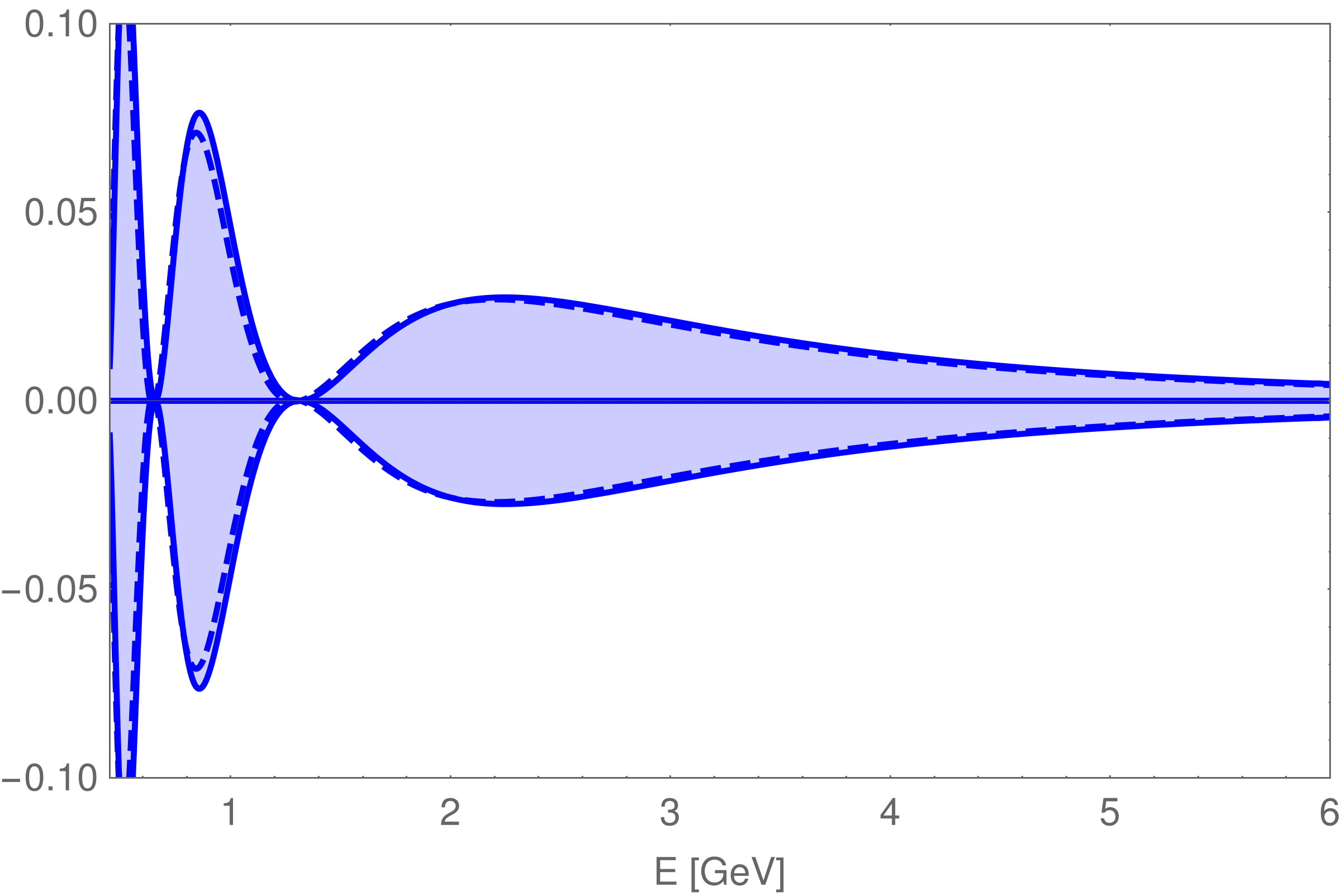}
		};

		\node at (2.35,1.5)[below left]{DUNE};
	\end{tikzpicture}
	\caption{Energy distribution of the T-odd component of $\asym{CP}_{\mu e}$
	at T2HK $L = 295$~km (left) and DUNE $L = 1300$~km (right), which is Hierarchy independent.
	Both the exact (dashed) and the analytical (solid) results from Eq.~(\ref{eq:analAT}) are shown.
	The bands correspond to all possible values changing $\delta$ in $(0,\,2\pi)$;
	the upper/central/lower lines correspond to $\sin\delta = -1,\, 0,\, 1$.
	Notice the different energy range and scale of the asymmetry between T2HK and DUNE plots.}
	\label{fig:AT}
\end{figure}

The decreasing value of the $\mathcal{\tilde J} / \mathcal{J}$ ratio 
with higher energies described by Eqs. (\ref{eqs:analImJ_highE}), 
i.e. when $\abs{a} \gg \dm_{21}$, is a consequence of the
absence of genuine CP violation in matter in the limit $\dm_{21} = 0$ even if
there are three non-vanishing neutrino masses in matter. The transmutation
of masses in vacuum to mixings in matter forces the smallness of 
the imaginary part of the rephasing-invariant mixing in matter at high energies.
However, this fact does not necessarily mean that 
genuine CP violation is unobservable at these energies, 
since the genuine CPV component of the CP asymmetry contains
this energy-dependent factor together with the matter-dependent oscillation function 
---odd in $L$--- that depends on both energy and baseline.
The effects of the baseline are shown in Fig.~\ref{fig:AT},
comparing the whole $\asym{T}_{\mu e}$ 
as function of the energy for
T2HK $L = 295$~km and DUNE $L=1300$~km,
where it is seen that the oscillation amplitude of each of them 
(fixed $L$) decreases as $1/E$, as expected,
and a higher baseline (at fixed $E$) enhances the values of $\asym{T}_{\mu e}$.

This behavior is understood
with the perturbation expansion in $\abs{U_{e3}}^2 \ll 1$ in the
energy regime between the two MSW resonances,
$\dm_{21} \ll \abs{a} \ll \abs{\dm_{31}}$, that we performed in the previous
Section.
The $1/E$ dependence in $\mathcal{\tilde J}$ is changed
by the approximated oscillating functions into $L/E$,
producing genuine CPV components of the same size
at the spectrum peak of both experiments.
In fact, the matter effects in $\mathcal{\tilde J}$
and oscillating phases just compensate to generate in this approximation
a genuine CPV asymmetry equal to that in vacuum, i.e. Eq.~(\ref{eq:analAT}). 
As such, it is odd in $L/E$, independent of $a$  and the Hierarchy, and
proportional to $\sin\delta$.

\section{Neutrino mass ordering discrimination}
\label{sec:Hier}

Last Section has demonstrated that the genuine $\asym{T}_{\mu e}$ component
of the CP asymmetry in matter is, to a good approximation 
for energies ---as planned in accelerator facilities--- between the two resonances 
$\dm_{21} \ll \abs{a} \ll \abs{\dm_{31}}$,
given by the vacuum CP asymmetry.
Its information content is then crucial to identify experimental
signatures of genuine CPV. On the other hand, it has nothing to say about 
the neutrino mass ordering: it is invariant under the change of sign in
$\dm_{31}$.
This simple change of sign, without changing the absolute value $\abs{\dm_{31}}$, 
is in fact the only effect of changing the hierarchy under the approximations
leading to Eqs.~(\ref{eqs:analAsyms}).

This Section discusses the information on the neutrino mass ordering 
contained in $\asym{CPT}_{\mu e}$, which is even in $L$ and $\sin\delta$
and odd in $a$. 
Propagation in matter is needed to generate effects of the change of hierarchy and
our $\asym{CPT}_{\mu e}$ is able to separate out this information, 
going beyond studies of its influence on transition
probabilities~\cite{shao-feng}.

There is no simple matter-vacuum relation such as Eq.~(\ref{eq:invariantImJ})
to easily write Re$\tilde J_{\alpha\beta}^{ij}$ as function of the vacuum
Re$J_{\alpha\beta}^{ij}$ ---the most compact result following this idea 
is~\cite{invarRe_kimura, invarRe_scott}
\begin{equation}
	\dmt_{12}\dmt_{23}\dmt_{31}\dmt_{ij}\,
	\mathrm{Re} \tilde J_{\alpha\beta}^{ij}
	= K_{\alpha\beta}^{ij} +
	\dm_{12}\dm_{23}\dm_{31}\dm_{ij}\,
	\mathrm{Re} J_{\alpha\beta}^{ij}\,,
\end{equation}
where all $K_{\alpha\beta}^{ij}$ vanish in vacuum.
This relation explains the dependence of all $L$-even terms in the oscillation
probabilities in each of the $\tilde \Delta_{ij}$ phases as 
$\frac{1}{\tilde \Delta_{ij}^2} \sin^2\tilde \Delta_{ij}$,
which is the reason why the vacuum limit $a\to 0$ is restored in these observables even
after taking $\dm_{21}\ll\abs{a}$, as discussed in
Section~\ref{sec:desarrollos}.
However, the $K_{\alpha\beta}^{ij}$ are complicated functions of the vacuum
quantities, and do not provide a clear insight into the behavior of
$\asym{CPT}_{\alpha\beta}$, so we will use Eq.~(\ref{eq:analACPT}) instead.

In general, this matter-induced component of the CP asymmetry 
has no definite transformation properties under the change of sign in $\dm_{31}$. 
Under the approximations made in Section~\ref{sec:desarrollos}, 
there are two distinct terms in $\asym{CPT}_{\mu e}$, a first one $\asym{CPT}_{-}$ 
which is an odd function of $\dm_{31}$ and a second one $\asym{CPT}_{+}$ 
which is an even function of $\dm_{31}$,
%\begin{subequations}
%	\label{eqs:ACPTpm}
%	\begin{align}
%		\nonumber
%		\asym{CPT}_{\mu e} &= \asym{CPT}_{-} + \asym{CPT}_{+}\,,\\
%		\label{eq:ACPT-}
%		&\asym{CPT}_{-} =
%		16\,S A\sin\Delta_{31}
%			\left[ \frac{\sin\Delta_{31}}{\Delta_{31}}-\cos\Delta_{31} \right]
%			\,,\\
%		\label{eq:ACPT+}
%		&\asym{CPT}_{+} = 
%			16\, J_r \cos\delta\, A\, \Delta_{21}\cos\Delta_{31}
%			\left[ \frac{\sin\Delta_{31}}{\Delta_{31}}-\cos\Delta_{31} \right]
%			\,.
%	\end{align}
%\end{subequations}
\begin{subequations}
	\label{eqs:ACPTpm}
	\begin{align}
		\nonumber
		\asym{CPT}_{\mu e} &= \asym{CPT}_{-} + \asym{CPT}_{+}\,,\\
		\label{eq:ACPT-}
		&\asym{CPT}_{-} =
		16\,A
			\left[ \frac{\sin\Delta_{31}}{\Delta_{31}}-\cos\Delta_{31} \right]
		S\sin\Delta_{31}
			\,,\\
		\label{eq:ACPT+}
		&\asym{CPT}_{+} = 
			16\, A
			\left[ \frac{\sin\Delta_{31}}{\Delta_{31}}-\cos\Delta_{31} \right]
			J_r \cos\delta\, \Delta_{21}\cos\Delta_{31}
			\,.
	\end{align}
\end{subequations}
Notice that both terms, as well as the whole $\asym{CPT}_{\mu e}$, vanish 
simultaneously when the $\delta$-independent common prefactor vanishes.
Alternatively, $\asym{CPT}_{\mu e}$ vanishes $\delta$-dependently when these
$\asym{CPT}_{-}$ and $\asym{CPT}_{+}$ terms compensate each other.

\begin{figure}[b!]
	\centering
	\begin{tikzpicture}[line width=1 pt, scale=1.5]
		\node at (0,0){
			\includegraphics[width=0.475\textwidth]{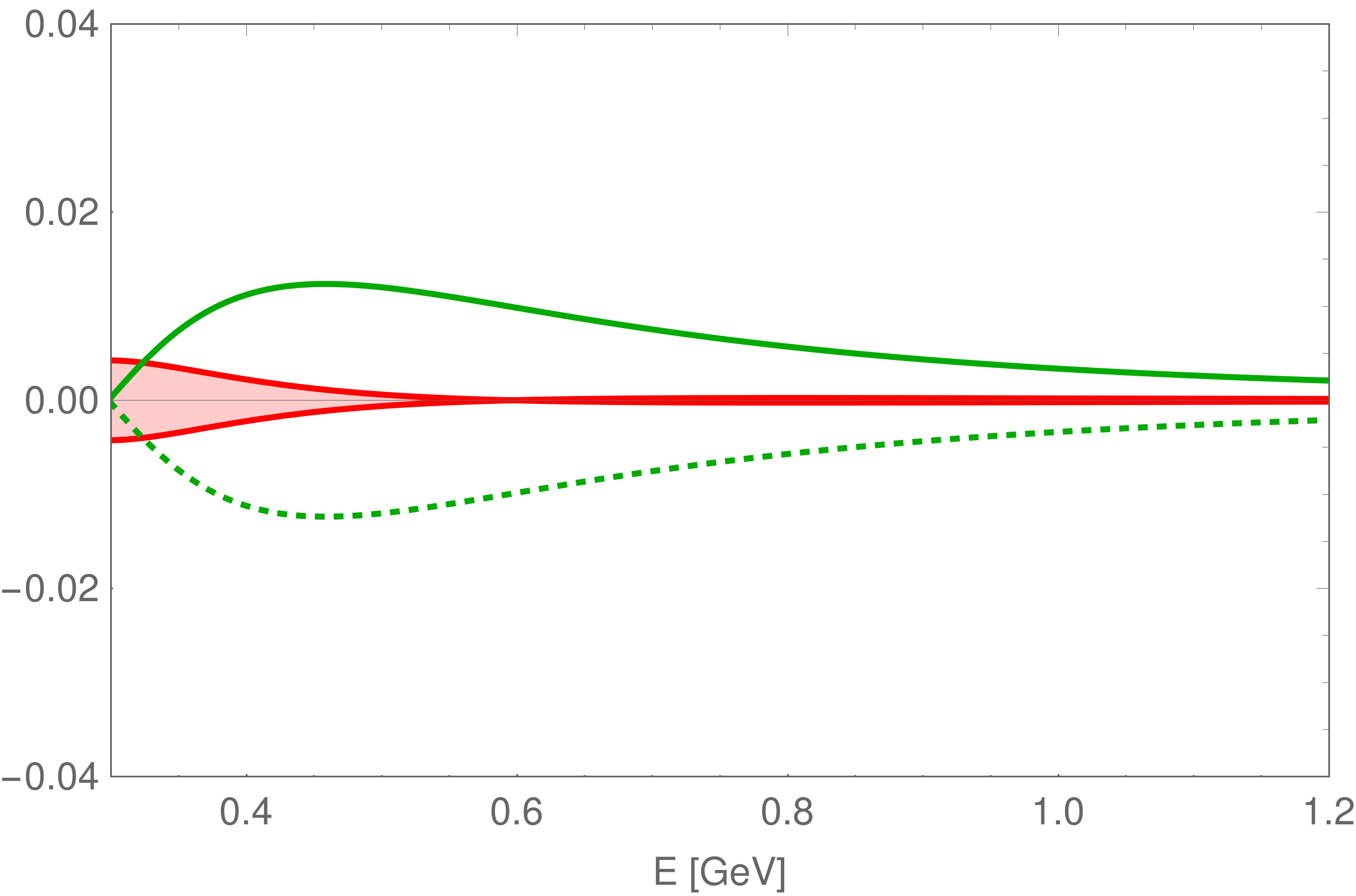}
		};

		\node at (2.35,1.5)[below left]{T2HK};
	\end{tikzpicture}
	\hfill
	\begin{tikzpicture}[line width=1 pt, scale=1.5]
		\node at (0,0){
			\includegraphics[width=0.475\textwidth]{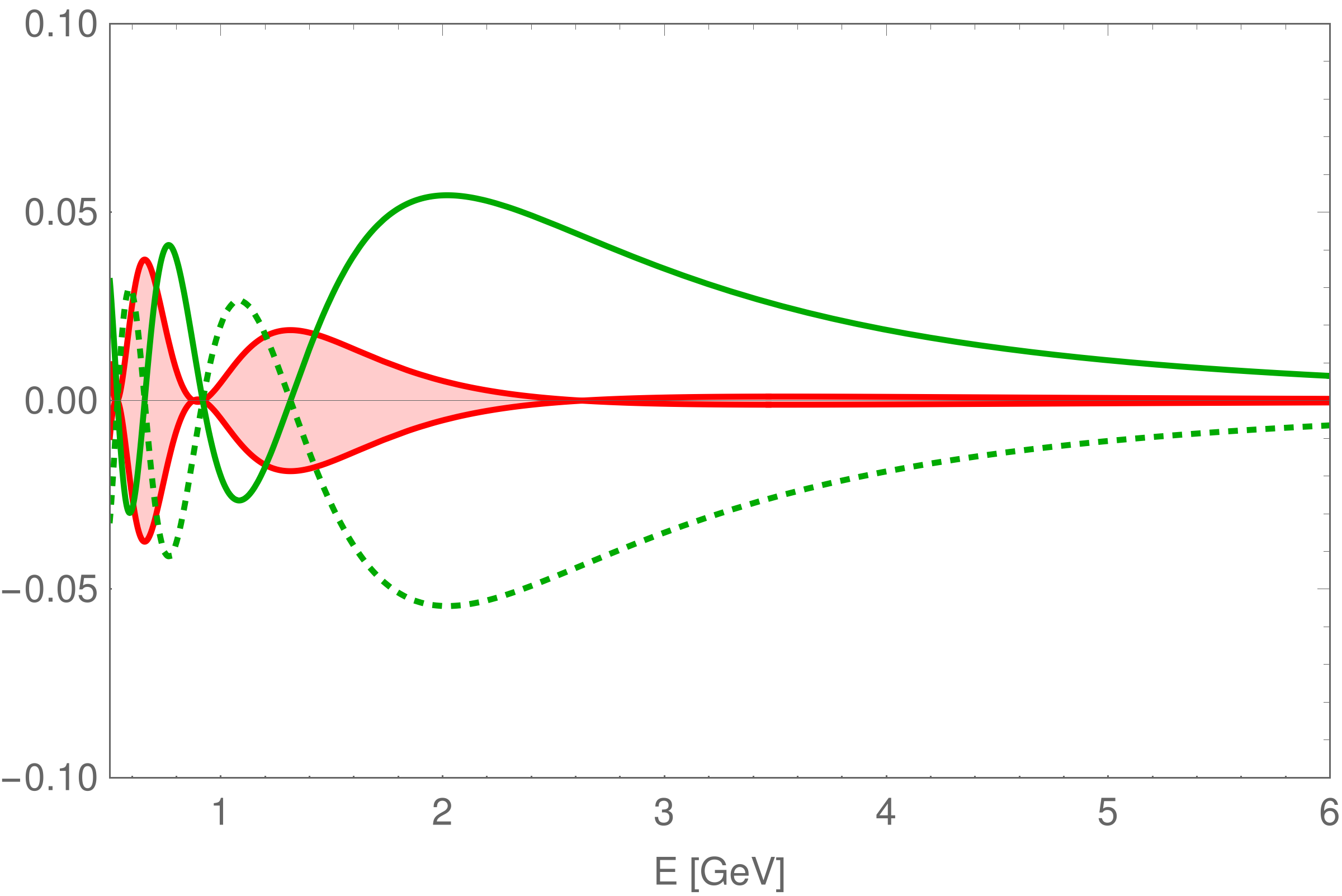}
		};

		\node at (2.35,1.5)[below left]{DUNE};
	\end{tikzpicture}
	\caption{Energy distribution of the two distinct terms of $\asym{CPT}_{\mu e}$
	as defined in Eqs.~(\ref{eqs:ACPTpm}),
	$\asym{CPT}_-$~(green, $\delta$-independent, hierarchy-odd) and 
	$\asym{CPT}_+$ (red, $\cos\delta$-odd, hierarchy-invariant),
	at T2HK $L = 295$~km (left) and 
	DUNE $L = 1300$~km (right).
	Both Normal Hierarchy (solid) and Inverted Hierarchy (dashed) shown.
	The bands correspond to all possible values changing $\delta$ in $(0,\,2\pi)$. 
	Notice the different energy range and scale of the asymmetry between T2HK and DUNE plots.}
	\label{fig:ACPT}
\end{figure}

As seen, the information content in $\asym{CPT}_{\mu e}$ on the
neutrino mass hierarchy is due to $\asym{CPT}_-$, its dominant zeroth-order term
in $\dm_{21}$, independent of the phase $\delta$. In the limit $\dm_{21}\to0$,
our results from Eq.~(\ref{eq:masses_exp}) in Section~\ref{sec:actual_experiments} show that
the mass spectrum in matter changes under a change of hierarchy
from neutrinos to antineutrinos as
\begin{equation}
	\label{eq:mtildeH}
	\Delta \tilde m_{21}^2 \leftrightarrow \Delta \tilde {\bar m}_{21}^2\,,
	\hspace{1cm}
	\Delta \tilde m_{31}^2 \leftrightarrow -\Delta \tilde {\bar m}_{32}^2\,,
	\hspace{1cm}
	\Delta \tilde m_{32}^2 \leftrightarrow -\Delta \tilde {\bar m}_{31}^2\,,
\end{equation}
whereas the $\tilde J_{\alpha\beta}^{ij}$ do not change sign,
so all $L$-even terms in the oscillation probabilities
---which are blind to the sign change in Eq.~(\ref{eq:mtildeH})---
are simply interchanged between neutrinos and antineutrinos.
As the CP asymmetry is a difference between neutrino and antineutrino oscillation probabilities, 
we discover that $\asym{CPT}_{\mu e}$ is only changing its sign 
under a change of hierarchy in the vanishing limit of $\dm_{21}$.

The $\asym{CPT}_{+}$ term in Eqs.~(\ref{eqs:ACPTpm}) is appreciable only at low energies, needing a 
non-vanishing $\dm_{21}$ and then sensitive to the $\delta$ phase
as a CP conserving $\cos\delta$ factor. In Fig.~\ref{fig:ACPT} we represent
these two components of $\asym{CPT}_{\mu e}$ as function of $E$ for the baselines of T2HK and
DUNE for Normal and Inverted Hierarchies.
To test the neutrino mass ordering from $\asym{CPT}_{\mu e}$,
we find that imposing the condition 
$\abs{\asym{CPT}_{-}} > \abs{\asym{CPT}_{+}}$
in the non-oscillating (high energy) region
leads to $E > 1.1\, E_{1^\mathrm{st}\;\mathrm{node}}$.
For~these energies above
the first node of the vacuum oscillation probability,
the whole effect of the change of sign
in $\dm_{31}$ is an almost odd $\asym{CPT}_{\mu e}$.
In addition, the $\asym{CPT}_{-}$ term in $\asym{CPT}_{\mu e}$ dominates the whole CP asymmetry 
at long baselines, as seen in Fig.~\ref{fig:Asyms}, so the measurement of the sign of
$\asym{CP}_{\mu e}$ at these energies fixes the hierarchy.

\section{Signatures of the peculiar energy dependencies}
\label{sec:signatures}

In this Section we identify those aspects of the energy distribution
of the CP asymmetry that can offer an experimental signature for the separation 
of its genuine and matter-induced components. 
With experiments in which the fingerprint of the baseline dependence, 
$L\text{-odd}$ and $L$-even functions, cannot be used, 
the peculiar patterns of the energy distribution provide precious information. 
The general trend of this dependence for $L = 1300$~km is given in Fig.~\ref{fig:Asyms}, 
showing the appearance of oscillations in the low and medium energy regions of the spectrum
with different behavior for the two components $\asym{T}_{\mu e}$ and
$\asym{CPT}_{\mu e}$, where nodes and extremal values are at different energies. 

\begin{figure}[b!]
	\centering
	\begin{tikzpicture}[line width=1 pt, scale=1.5]
		\node at (0,0){
			\includegraphics[width=0.475\textwidth]{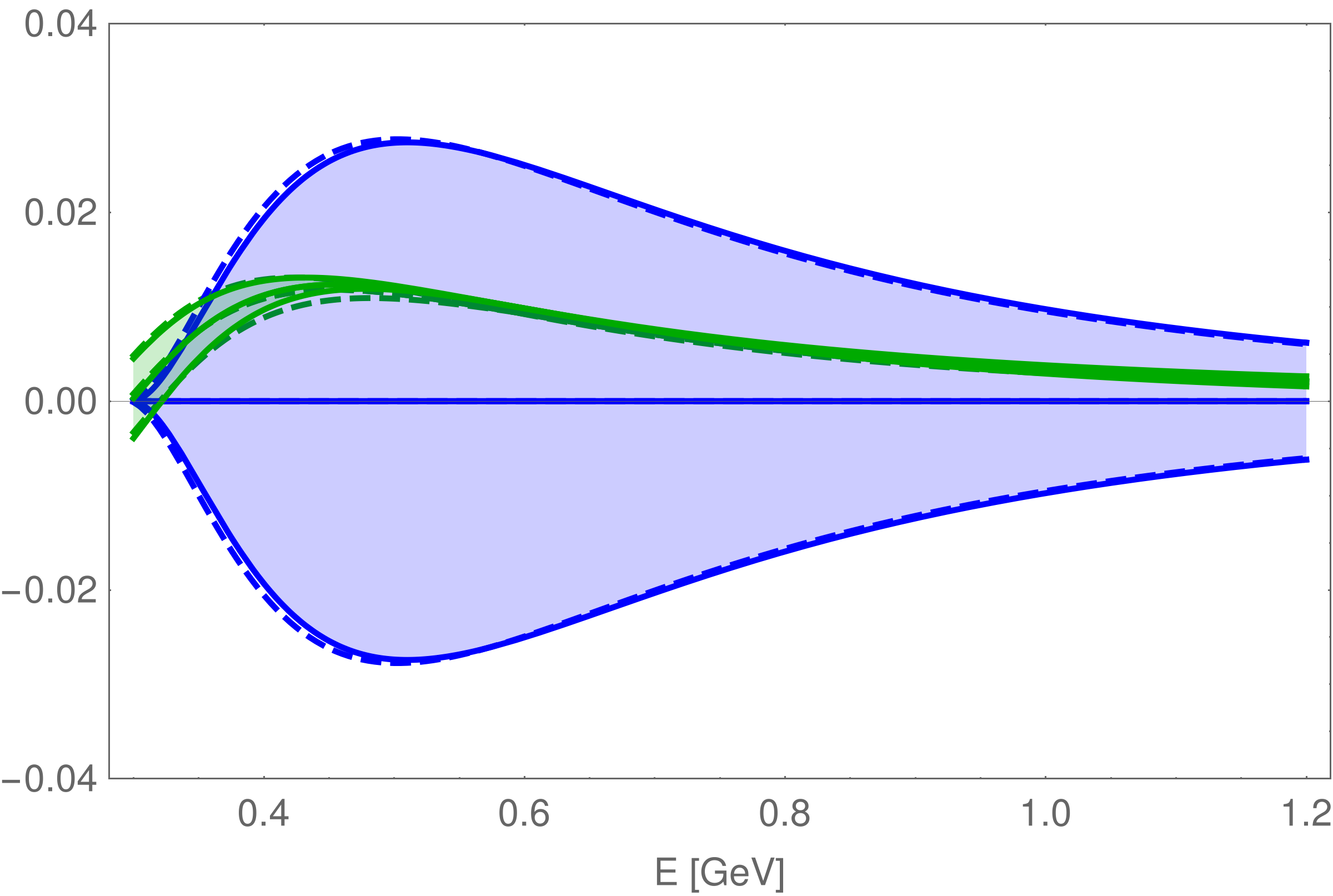}
		};

		\node at (2.4,1.5)[below left]{NH};
	\end{tikzpicture}
	\hfill
	\begin{tikzpicture}[line width=1 pt, scale=1.5]
		\node at (0,0){
			\includegraphics[width=0.475\textwidth]{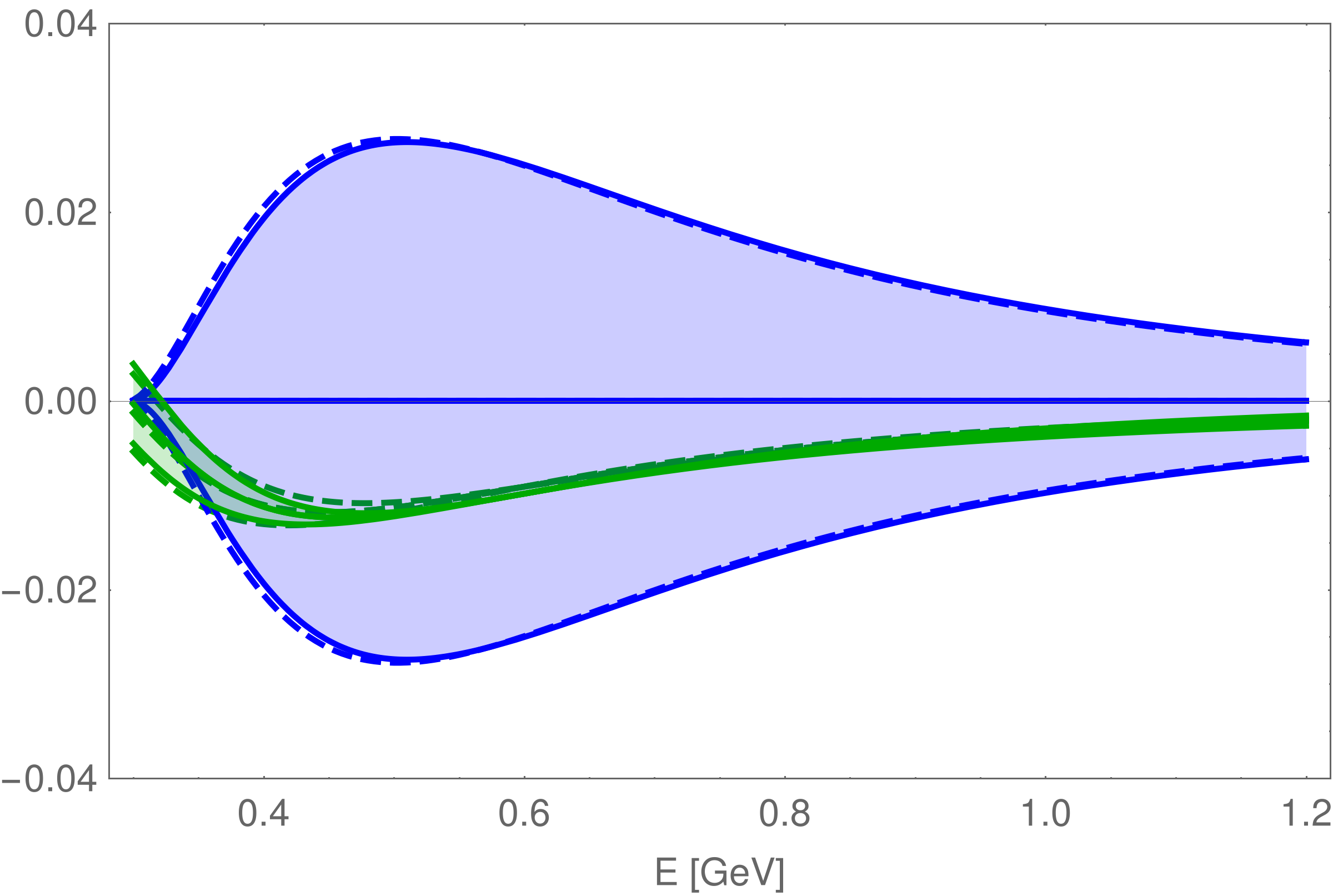}
		};

		\node at (2.4,1.5)[below left]{IH};
	\end{tikzpicture}
	\caption{CPT-odd (green) and T-odd (blue) components of $\asym{CP}_{\mu e}$
	as functions of the neutrino energy $E$ through the Earth mantle (of constant density)
	at T2HK baseline $L = 295$~km.
	Both the exact (dashed) and the analytical (solid)
	results from Eqs.(\ref{eqs:analAsyms}) are shown.
	Normal/Inverted Hierarchy in the left/right pannel.
	The bands correspond to all possible values changing $\delta$ in $(0,\,2\pi)$;
	the upper/central/lower lines for $\asym{CPT}_{\mu e} (\asym{T}_{\mu e})$ 
	correspond to $\cos\delta (\sin\delta) = -1,\, 0,\, 1$.
	}
	\label{fig:AsymsHK}
\end{figure}

However, this rich structure is lost when the baseline is decreased to
$L=295$~km and a threshold energy of $300$~MeV is imposed. 
The emerging picture under these conditions is given in Fig.~\ref{fig:AsymsHK} 
and the main conclusion is the relative suppression of $\asym{CPT}_{\mu e}$
with respect to $\asym{T}_{\mu e}$, due to its proportionality to $A \propto L$.
In addition, this small $\asym{CPT}_{\mu e}$ is mainly the $\delta$-independent 
$\asym{CPT}_{-}$ in Eq.~(\ref{eq:ACPT-}) and so it can be subtracted away from the 
experimental $\asym{CP}_{\mu e}$, if the neutrino mass hierarchy is previously known, 
as a theoretical background. This would allow to separate the genuine
$\asym{T}_{\mu e}$ component.

Using the analytical approximate expressions of the observable components 
given in Eqs.~(\ref{eqs:analAsyms}), we perform a detailed study of the
position of extremal values and zeros of each of them, 
as well as their behavior around the zeros. 
The energy dependence is controlled by the phase $\Delta_{31} \propto 1/E$ 
and we will take as reference the functional form of the CP-conserving transition 
probability $f(\Delta) = \sin^2\Delta$.

\begin{figure}[b!]
	\centering
	\includegraphics[width=0.7\textwidth]{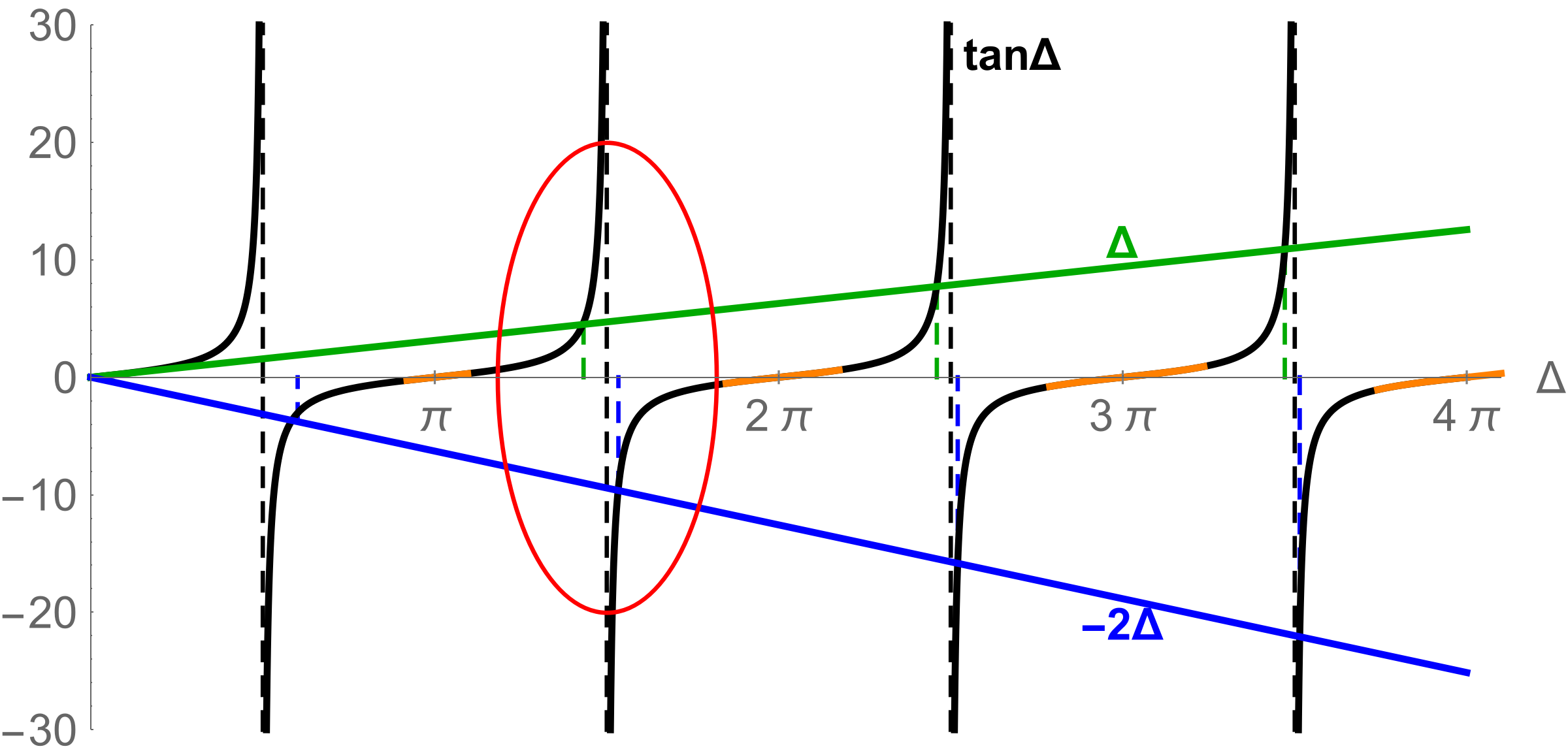}
	\caption{
		Illustration of the position of the relevant zeros of $\asym{CPT}_{\mu e}$,
		given by $\tan\Delta_{31} = \Delta_{31}$, 
		and the maxima of $\abs{\asym{T}_{\mu e}}$,
		given by $\tan\Delta_{31} = -2\Delta_{31}$.
		The vertical dashed lines are the asymptotes of $\tan\Delta$ (black),
		corresponding to oscillation maxima,
		and the perturbative solutions of the previous equations,
		$\Delta_0^\mathrm{CPT} = (2n+1)\frac{\pi}{2} - \left[(2n+1)\frac{\pi}{2}\right]^{-1}$
		(green) and
		$\Delta_\mathrm{max}^\mathrm{T} = (2n+1)\frac{\pi}{2} + \frac{1}{2}\left[(2n+1)\frac{\pi}{2}\right]^{-1}$
		(blue).
		As calculated in Eq.~(\ref{eq:CPT0delta}), 
		the $\delta$-dependent zeros of $\asym{CPT}_{\mu e}$ around $\Delta_0=n\pi$
		are bounded within the orange regions of $\tan \Delta$.
		The highest-energy point, i.e. smallest $\Delta$, where
		$\asym{CPT}_{\mu e}$ vanishes $\delta$-independently is emphasized by the red ellipse.
	}
	\label{fig:tanx}
\end{figure}

For the genuine component $\asym{T}_{\mu e}$, the energy distribution is
\begin{equation}
	\label{eq:fT}
	\f{T}(\Delta) = \Delta \sin^2\Delta \,.
\end{equation}
Contrary to $f(\Delta)$, the amplitude of the oscillations of 
$\f{T}(\Delta)$ decreases as $1/E$, 
but the zeros are the same $\Delta_0 = 0,\, \pi,\, 2\pi \ldots$ as for $f(\Delta)$. 
There are, however, two series of extremal values. 
The first kind, those corresponding to solutions
of $\sin\Delta = 0$ as for the zeros,
are double zeros, which
indicates that $\asym{T}_{\mu e}$ keeps the same
sign around the zeros,
and so in the whole energy spectrum. 
From Eq.~(\ref{eq:analAT}) it is clear that the sign is given by
$\mathrm{sign}(\asym{T}_{\mu e}) = -\mathrm{sign}(\sin\delta)$.

The additional extremal values appear for
\begin{equation}
	\label{eq:tanx2x}
	\tan\Delta +2\Delta = 0 \,.
\end{equation}
In Fig~\ref{fig:tanx} we identify graphically the solutions to Eq.~(\ref{eq:tanx2x}), 
which appear slightly above the oscillation maxima
$\Delta_\mathrm{max}^\mathrm{osc} = (2n+1)\frac{\pi}{2}$.
This is a first fortunate fact, implying that the experimental configurations
with maximal $\abs{\asym{T}_{\mu e}}$ are close to those with highest
statistics.
A perturbative expansion of $\cot\Delta$ around
$\Delta_\mathrm{max}^\mathrm{osc}$ leads to the approximate solutions
\begin{align}
	\nonumber
	\Delta_\mathrm{max}^\mathrm{T} &=
	(2n+1)\frac{\pi}{2} + \frac{1}{2}\left[(2n+1)\frac{\pi}{2}\right]^{-1}
	+\, \cdots\;, \hspace{1cm} n\geq 0 \\
	\label{eq:xTmax}
	&\approx
	\frac{\pi}{2} + \frac{1}{\pi},\; 
	\frac{3\pi}{2} + \frac{1}{3\pi},\; 
	\frac{5\pi}{2} + \frac{1}{5\pi}\ldots \,,
\end{align}
which show that the interesting (see below) second and higher maxima 
in $\abs{\asym{T}_{\mu e}}$ are
within a $3\%$ interval above the oscillation maxima.

In the case of the matter-induced $\asym{CPT}_{\mu e}$ component of the CP asymmetry,
the energy distribution in Eq.~(\ref{eq:analACPT}) is
\begin{equation}
	\label{eq:fCPT}
	\f{CPT} (\Delta) = 
	\left( \frac{\sin\Delta}{\Delta}-\cos\Delta \right)
	\left( S\sin\Delta + J_r\, \frac{\dm_{21}}{\dm_{31}}\cos\delta\,\Delta\cos\Delta \right)
	\,,
\end{equation}
which has two kinds of zeros with distinct implications.
The vanishing condition for the second factor are the $\delta$-dependent
solutions of
\begin{equation}
	\label{eq:CPT0delta}
	\tan\Delta = -\frac{J_r}{S}\, \frac{\dm_{21}}{\dm_{31}}\cos\delta\; \Delta
		= -0.09\cos\delta\; \Delta\,,
\end{equation}
that reduce to vacuum nodes $\sin\Delta=0$ if $\cos\delta =0$,
where $\asym{T}_{\mu e}$ also vanishes.
The actual position of these zeros is strongly dependent on $\cos\delta$,
and the set of solutions is illustrated in Fig.~\ref{fig:tanx} 
by the region around $\Delta_0 = n\pi$ where $\tan\Delta$ is orange.

The second kind of zeros in Eq.~(\ref{eq:fCPT}) are solutions of the equation
\begin{equation}
	\tan\Delta = \Delta\,,
\end{equation}
and are graphically depicted in Fig.~\ref{fig:tanx} too. 
As seen, they appear slightly below the oscillation maxima in $f(\Delta)$ 
starting from the second one, with approximate values
\begin{align}
	\nonumber
	\Delta_0^\mathrm{CPT} &=
	(2n+1)\frac{\pi}{2} -\left[(2n+1)\frac{\pi}{2}\right]^{-1}
	+\, \cdots\;, \hspace{1cm} n\geq 1 \\
	\label{eq:xCPT-0}
	&\approx
	\frac{3\pi}{2} - \frac{2}{3\pi},\; 
	\frac{5\pi}{2} - \frac{2}{5\pi}\ldots \,,
\end{align}
which almost coincide with the maxima 
$\Delta_\mathrm{max}^\mathrm{T}$ in Eq.~(\ref{eq:xTmax}) of $\abs{\asym{T}_{\mu e}}$.
Not only that: these zeros $\Delta_0^\mathrm{CPT}$ of $\asym{CPT}_{\mu e}$ are again 
near the oscillation maxima 
$\Delta_\mathrm{max}^\mathrm{osc}=(2n+1)\frac{\pi}{2}$, 
so we conclude that there are 
``magic energies'' at these (\ref{eq:xCPT-0}) phase values,
within a $5\%$ interval below the corresponding oscillation maximum, in which 
$\asym{CPT}_{\mu e}$ vanishes and $\abs{\asym{T}_{\mu e}}$ is close to a
maximum.
These magic points have additional bonuses: i) the zero of $\asym{CPT}_{\mu e}$
is independent of $\cos\delta$, providing no ambiguity in its position;
ii) these are simple zeros, in such a way that the sign of $\asym{CPT}_{\mu e}$
is changing around them;
iii) although $\abs{\asym{T}_{\mu e}}$ is not exactly at its maximum value
when $\asym{CPT}_{\mu e}=0$,
the leading order deviations from $\Delta_\mathrm{max}^\mathrm{osc}$ we
calculated show that its value is above 
$90\% \abs{\asym{T}_{\mu e}}_\mathrm{max}$.
A look into the derivative of $\f{CPT}(\Delta)$ shows that 
the sign-change of $\asym{CPT}_{\mu e}$ around these zeros is such that
$\asym{CPT}_{\mu e}$ 
is always decreasing (increasing) around the relevant $\delta$-independent
zeros for Normal (Inverted) Hierarchy, and opposite around $\delta$-dependent
zeros.

Taking into account the dependence in $L/E$
of these remarkable values of the phases, we give in Table~\ref{tab:MagicPoint} 
the relevant energies around the second oscillation maximum for both
the baselines of the T2HK and DUNE experiments. The precise position 
of this energy,
which is slightly above the second oscillation maximum,
%varies linearly with $\abs{\dm_{31}}$ and the baseline as
is proportional to $L\abs{\dm_{31}}$ as

\vspace{6pt}
\noindent
\boxed{
	\begin{minipage}{0.99\textwidth}
		\vspace{-0cm}
		\begin{equation}
		\label{eq:magicE}
			E = 0.92~\mathrm{GeV}\,\frac{L}{1300~\mathrm{km}}
			\,\frac{\abs{\dm_{31}}}{2.5\times 10^{-3}~\mathrm{eV}^2}\,,
		\end{equation}
		\vspace{-0.4cm}
	\end{minipage}
}

\vspace{6pt}
\noindent
which explains the absence of this rich oscillatory structure in Fig.~\ref{fig:AsymsHK}:
at the short baseline of T2HK, all interesting points lie below the threshold
energy of $300$~MeV.

\begin{table}[b!]
	\begin{center}
		\caption{
			Specific position of 
			the first zero $\Delta_0^\mathrm{CPT}$ in Eq.~(\ref{eq:xCPT-0}),
			the second vacuum oscillation maximum and the
			second maximum $\Delta_\mathrm{max}^\mathrm{T}$ in Eq.~(\ref{eq:xTmax}),
			corresponding to the 
			highest-energy zero of $\asym{CPT}_{\mu e}$ independent of $\delta$.
			For each of these three points,
			we show the value of the oscillation phase, 
			which is independent of any experimental parameter;
			the $L/E$, 
			whose value depends linearly on the inverse of $\abs{\dm_{31}}$;
			and the particular energy associated to this $L/E$
			for T2HK $L=295$~km and DUNE $L=1300$~km.
			Notice that these three values of the phase $\Delta_{31}$
			correspond to the position of the
			green/black/blue dashed lines within the red ellipse in Fig.~\ref{fig:tanx}.
		}
		\vspace{12pt}
		\label{tab:MagicPoint}
		\begin{tabular}{rcccc}
			\toprule
			&\multirow{2}{*}{$\Delta_{31}$}
			&\multirow{2}{*}{$\frac{L}{E}\left(\frac{\mathrm{km}}{\mathrm{GeV}}\right)$}
			&\multicolumn{2}{c}{$E$ (GeV)}\\ 
			\cline{4-5}
			&&&T2HK
			&DUNE\\
			\midrule
			Vanishing $\asym{CPT}_{\mu e}$
			\hspace{0.2cm}
				&4.50 &1420 &0.21 &0.92\\
			$2^\mathrm{nd}$ Oscillation Maximum
			\hspace{0.2cm}
				&4.71 &1480 &0.20 &0.88\\
			Maximum $\abs{\asym{T}_{\mu e}}$
			\hspace{0.2cm}
				&4.82 &1520 &0.19 &0.86\\
			\bottomrule
		\end{tabular}
	\end{center}
\end{table}

\begin{figure}[t!]
	\centering
	\begin{tikzpicture}[line width=1 pt, scale=1.5]
		\node at (0,0){
			\includegraphics[width=0.475\textwidth]{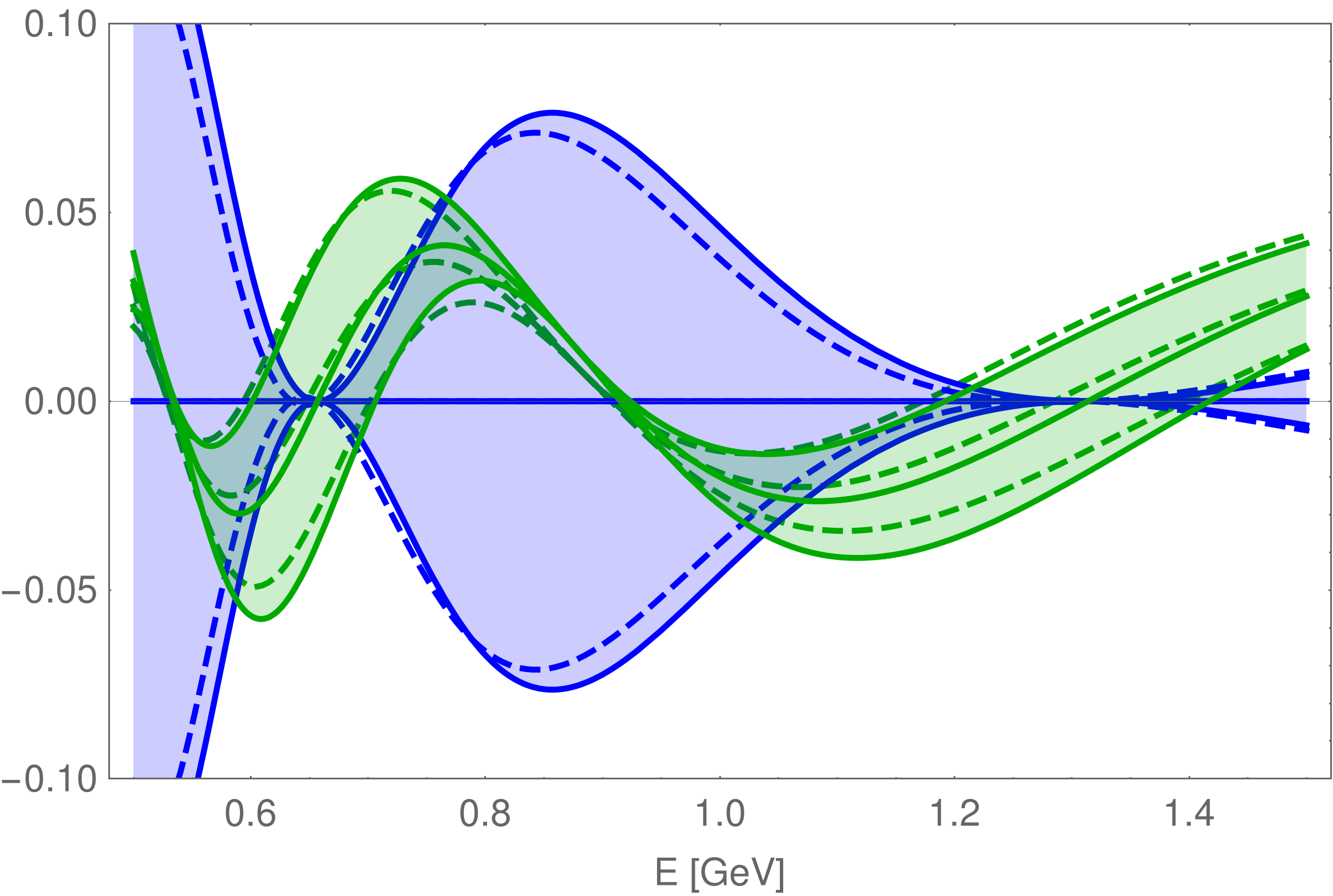}
		};

		\node at (2.4,1.5)[below left]{NH};
	\end{tikzpicture}
	\hfill
	\begin{tikzpicture}[line width=1 pt, scale=1.5]
		\node at (0,0){
			\includegraphics[width=0.475\textwidth]{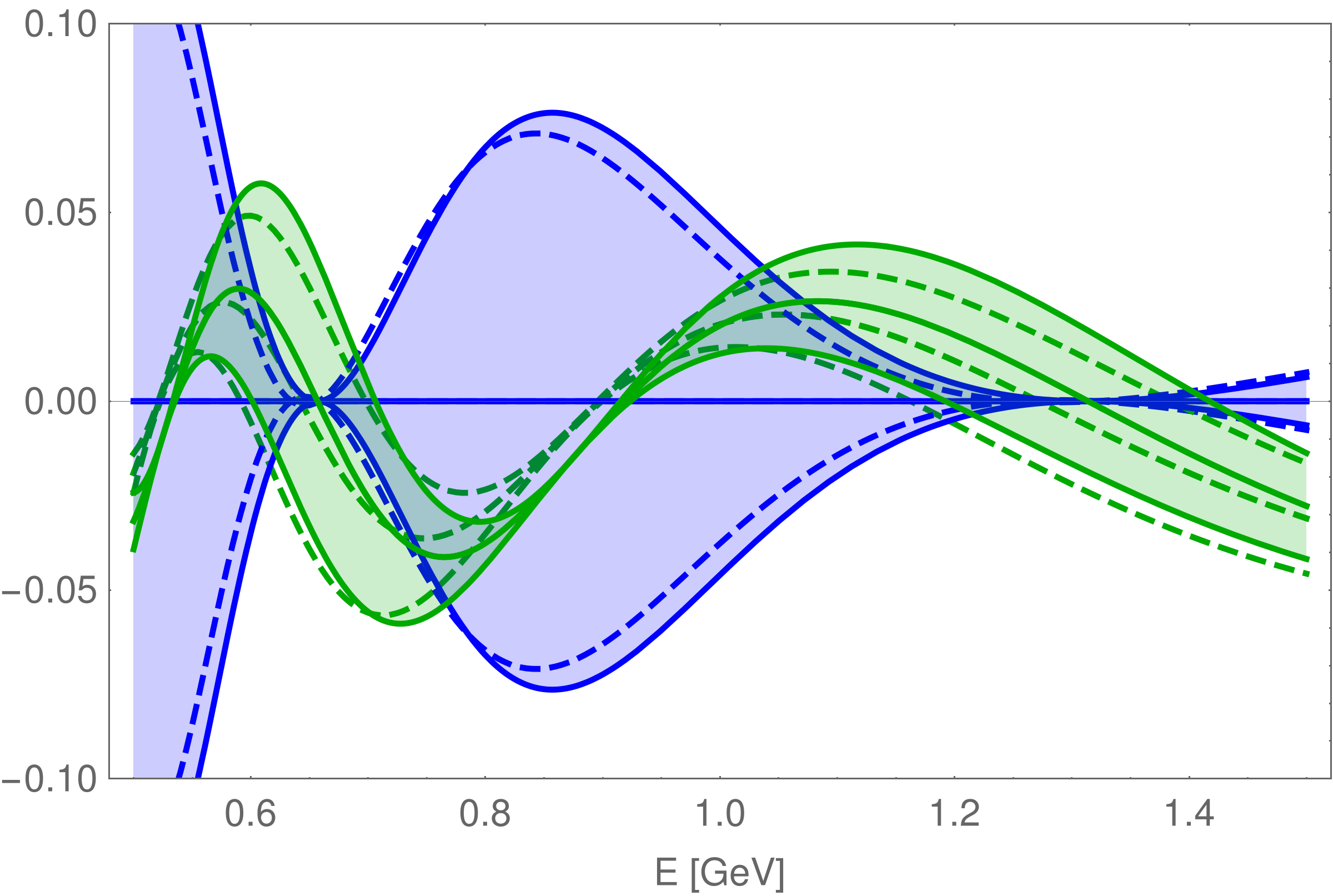}
		};

		\node at (2.4,1.5)[below left]{IH};
	\end{tikzpicture}
	\caption{Zoom of Fig.~\ref{fig:Asyms} at low $E$, showing
	$\asym{CPT}_{\mu e}$ (green) and $\asym{T}_{\mu e}$ (blue) 
	at DUNE $L = 1300$~km.
	%Normal Hierarchy in the left pannel, Inverted Hierarchy in the right pannel.
	}
	\label{fig:zoom}
\end{figure}

This magic configuration around the second oscillation maximum is well apparent 
in the results presented in Fig.~\ref{fig:zoom} for\footnote{ 
Notice that an equivalent Figure could be obtained at $L = 295$~km for energies
between 100 and 350~MeV, with the same energy-dependence for both components of
the CP asymmetry, but a relatively smaller $\asym{CPT}_{\mu e}$ due to its
proportionality to $A \propto L$.
}
$L = 1300$~km.
One can observe that
the uninteresting (increasing/decreasing for NH/IH) zeros in $\asym{CPT}_{\mu e}$ 
are strongly dependent on $\cos\delta$, as seen in Eq.~(\ref{eq:CPT0delta}), 
and their position when $\cos\delta=0$
is that of the $\delta$-independent zeros in $\asym{T}_{\mu e}$.
As understood from the previous discussion, 
we have identified the most relevant $\delta$-independent zeros (\ref{eq:xCPT-0}) of
$\asym{CPT}_{\mu e}$, 
decreasing/increasing for NH/IH,
correlated to near maximal $\abs{\asym{T}_{\mu e}}$ proportional to
$\sin\delta$.
%Since $\asym{CPT}_{\mu e}$ is changing its sign around the zero,
%integrating statistics in a bin around this point would still result in a
%vanishing matter term in the experimental CP asymmetry,
%providing a direct test of CP violation in the lepton sector as clean as in
%vacuum.
%We have checked this is the case for an energy bin width up to 
%$0.15-0.20$ GeV,
%which keeps an almost vanishing $\asym{CPT}_{\mu e}\,\forall \delta$ and an almost maximal 
%$\asym{T}_{\mu e} \propto \sin\delta$.
%Such an energy resolution seems to be feasible at DUNE~\cite{fdez-mtnez}.
Due to the first-order character of this $\delta$-independent
(and hierarchy-independent too) zero, 
$\asym{CPT}_{\mu e}$ is changing sign around it. 

\begin{figure}[t]
	\centering
	\begin{tikzpicture}[line width=1 pt, scale=1.5]
		\node at (0,0){
			\includegraphics[width=0.475\textwidth]{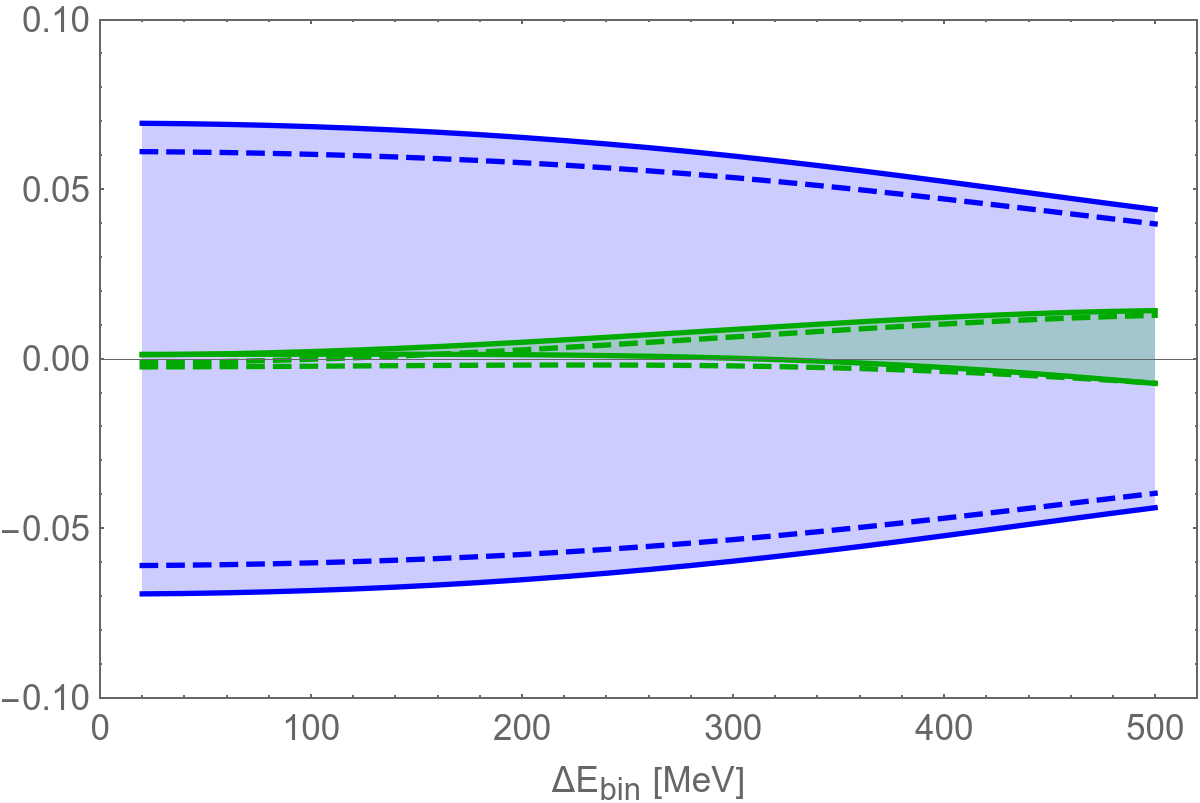}
		};

		\node at (2.4,1.5)[below left]{NH};
	\end{tikzpicture}
	\hfill
	\begin{tikzpicture}[line width=1 pt, scale=1.5]
		\node at (0,0){
			\includegraphics[width=0.475\textwidth]{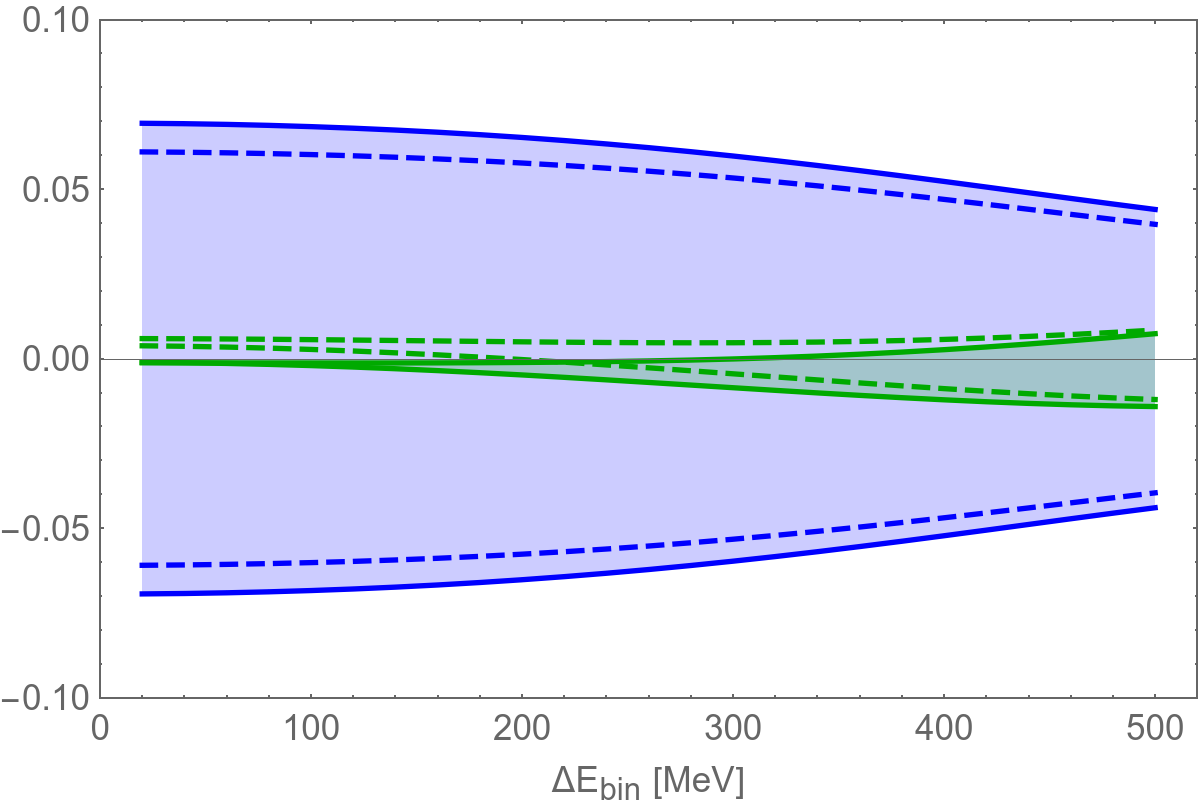}
		};

		\node at (2.4,1.5)[below left]{IH};
	\end{tikzpicture}
	\caption{
	    Average value of the CPT-odd (green) and T-odd (blue) compoments of
	    $\asym{CP}_{\mu e}$,
	    at DUNE baseline $L = 1300$~km,
	    in an energy bin width $\Delta E_\mathrm{bin}$
	    centered on the magic energy (\ref{eq:magicE}).
	    Both the exact (dashed) and the analytical (solid)
	    results from Eqs.(\ref{eqs:analAsyms}) are shown.
	    Normal/Inverted Hierarchy in the left/right pannel.
	    The bands correspond to all possible values changing $\delta$ in $(0,\,2\pi)$;
	    the upper/lower lines for $\asym{CPT}_{\mu e} (\asym{T}_{\mu e})$ 
	    correspond to $\cos\delta (\sin\delta) = -1,\, 1$.
	}
	\label{fig:bin}
\end{figure}

Integrating statistics in an energy bin around this point would still result 
in a vanishing fake CPV term in the experimental CP asymmetry, 
providing a direct test of CP violation in the lepton sector as clean as in vacuum. 
As shown in Fig.~\ref{fig:bin}, we have checked that this is the case for an energy bin width 
up to $0.15 - 0.20$ GeV,
which keeps an almost vanishing $\asym{CPT}_{\mu e}\,\forall \delta$ and an almost maximal 
$\asym{T}_{\mu e} \propto \sin\delta$.
Such an energy resolution appears to be feasible at DUNE~\cite{fdez-mtnez} around the second
oscillation maximum, and the accumulated events would provide enough
statistical significance to the transition probability distribution.

The whole discussion in this Section, 
which stems from the analytical expressions (\ref{eqs:analAsyms}), 
allows the reader to understand the peculiar energy distributions of 
the two components of the experimental CP asymmetry.
In particular, the value of the magic energy (\ref{eq:magicE}), 
as well as its (in)dependence on the different oscillation parameters, is explained.
This result in the energy distribution of the experimental CP asymmetry
provides a positive response to our search of observable signatures able to separate out
the genuine and matter-induced components.

\section{Conclusions}
\label{sec:conclusions}

A direct evidence of genuine CP violation means the measurement of
an observable odd under the symmetry. 
The CP asymmetry for long baseline neutrino oscillation experiments 
suffers from fake effects 
induced by the interaction with matter. This matter effect is, however, 
welcome as a source of information for the ordering of the neutrino mass spectrum. 
Based on the different transformation properties under T and CPT 
of the genuine and matter-induced CP violation 
we have proved a Disentanglement Theorem for these two components. 
In order to raise this disentanglement to a phenomenological separation 
of the two components we have identified in this work their peculiar signatures 
from a detailed study in terms of the experimentally accessible variables.

For a precise-enough understanding of the problem,
we have developed a new analytical perturbative expansion in both 
$\dm_{21},\, \abs{a}\ll\abs{\dm_{31}}$
without any assumption between $\dm_{21}$ and $a$, 
which we use to analyze each of the disentangled components of the CP asymmetry,
$\asym{CP}_{\alpha\beta} = \asym{CPT}_{\alpha\beta}+\asym{T}_{\alpha\beta}$,
the first one ($L$-even) accounting for matter effects,
the second one ($L$-odd) being genuine.

The two components of the CP violation asymmetry for the $\nu_\mu \to \nu_e$
transition are shown in Fig.~\ref{fig:Asyms_vacuum} as function of the
interaction parameter $a$. 
They fulfill all the T and CPT symmetry requirements 
proved in Section~\ref{sec:theorem}:
the CPT-odd component
$\asym{CPT}_{\alpha\beta}$ is an odd function of $a$ and vanishes linearly in
the limit $a\to 0$ for any value of the CP phase $\delta$, 
as well as being an even function of $\sin\delta$ due to T-invariance.
The T-odd component
$\asym{T}_{\alpha\beta}$ is an odd function of $\sin\delta$
that vanishes, even in matter, if there is no genuine CP violation,
as well as being even in $a$ due to CPT-invariance, which means that its value
is that of the CP asymmetry in vacuum up to small quadratic corrections
$\mathcal{O}(a/\dm_{31})^2$.

By analyzing the vacuum limit $a\to 0$ both above and below the T-invariant
limit $\dm_{21}\to 0$,
some intricacies for the mixings in matter appear. 
If one assumes $\abs{a}\ll\dm_{21}$, the vacuum limit of the mixing matrix in
matter will be the free PMNS matrix. On the other hand,
$\dm_{21} \ll \abs{a}$ will force $U_{e1}=0$ in the vacuum limit.
This different behavior stems from the fact that 
setting $\dm_{21}=0$ in the vacuum Hamiltonian leads to degenerate $\nu_1$,
$\nu_2$ mass eigenstates. The two limits mentioned above correspond to breaking
this degeneracy in favor of $\dm_{21}$ or $a$, respectively, projecting onto
different bases in the 12 subspace.
At the level of oscillation probabilities and asymmetries, the matter-vacuum invariant
relations studied in Sections~\ref{sec:AT} and \ref{sec:Hier},
which involve both mixings and masses, show that the
dependence on the phases associated to the small quantities 
$\epsilon = a,\, \dm_{21}$
are of the form $\frac{1}{\epsilon}\sin \frac{\epsilon L}{4E}$, which cancel out if both
of them are small, independently of whether $\abs{a}\ll \dm_{21}$ or
$\dm_{21}\ll \abs{a}$. Therefore, the commutability of the two limits $a\to 0$
and $\dm_{21}\to 0$ is restored for the final observables.

We have searched for experimental signatures in the $\nu_\mu \to \nu_e$ oscillation channel assuming
$\dm_{21} \ll \abs{a}$, valid for actual accelerator neutrino energies through the
Earth mantle. The definite $a$-parity of each component of the CP asymmetry
allows us to expand in $\abs{a} \ll \abs{\dm_{31}}$ to leading (linear in
$\asym{CPT}_{\mu e}$, constant in $\asym{T}_{\mu e}$) order, since corrections
are quadratic. 
For baselines and energies such that both $\epsilon = \dm_{21},\, a$ lead to
$\epsilon L/4E \ll 1$, 
and taking $\abs{U_{e3}}\ll 1$, we find compact expressions that faithfully
reproduce the exact results,

\begin{align*}
	\asym{CPT}_{\mu e} &=
	   16\, A 
	   \left[ \frac{\sin\Delta_{31}}{\Delta_{31}}-\cos\Delta_{31} \right]
	   \left( S\sin\Delta_{31} + J_r\cos\delta\, \Delta_{21}\cos\Delta_{31} \right)
	   +\mathcal{O}(A^3)\,,\\
	\asym{T}_{\mu e} &=
	   -16\,J_r \sin\delta\, \Delta_{21}\sin^2\Delta_{31}
	   +\mathcal{O}(A^2) \,,
\end{align*}
where $S \equiv c_{13}^2 s_{13}^2 s_{23}^2$,
$J_r \equiv c_{12} c_{13}^2 c_{23} s_{12} s_{13} s_{23}$,
$A\equiv \frac{aL}{4E} \propto L$ and
the two $\Delta_{ij} \equiv \frac{\dm_{ij} L}{4E} \propto~\!L/E$.
Equipped with such precise-enough analytical results, 
we have performed a detailed study of the different features of these
quantities, focusing especially on signatures of genuine CP violation and
hierarchy effects.

Since $\asym{T}_{\mu e}$ is blind to sign$(\dm_{31})$, a determination of the
neutrino mass ordering must come from regions where the hierarchy-odd (and
$\delta$-independent) term of $\asym{CPT}_{\mu e}$ dominates,
which can only happen at long baselines
due to the proportionality of $\asym{CPT}_{\mu e}$ to $A\propto L$.
Our analysis at DUNE $L=1300$~km shows that
this is the case for energies above the first node of the vacuum
oscillation, where the sign of the experimental $\asym{CP}_{\mu e}$ determines
the hierarchy.

The strategy towards the measurement of genuine CP violation depends on the
baseline. 
At medium baselines such as T2HK $L=295$~km, 
the CPT-odd component $\asym{CPT}_{\mu e}$ is small and,
for energies above the first oscillation node, dominated by its
$\delta$-independent term. Therefore, it can be theoretically subtracted from
the experimental $\asym{CP}_{\mu e}$, if the hierarchy is previously known, in
order to obtain the genuine component $\asym{T}_{\mu e}$.

At long baselines, both $\asym{CPT}_{\mu e}$ and $\asym{T}_{\mu e}$ are of the
same order, so
$\asym{CP}_{\mu e}$ will directly test genuine CP violation 
only when the CPT-odd component vanishes.
We find a family of simple zeros of $\asym{CPT}_{\mu e}$ with
decreasing/increasing slope for Normal/Inverted Hierarchy 
corresponding to the solutions of $\tan\Delta_{31}=\Delta_{31}$. 
These zeros are close to the second and higher vacuum oscillation maxima 
$\sin^2\Delta_{31} = 1$, implying that
their position is independent of $\delta$ and corresponds to a nearly maximal
$\abs{\asym{T}_{\mu e}}$ proportional to $\sin\delta$.

The main conclusion is thus that the magic energy around the second oscillation
maximum is the ideal choice to find a direct evidence of
genuine CP violation in the lepton sector.
This vanishing of $\asym{CPT}_{\mu e}$ occurs at
\begin{equation*}
	E = 0.92~\mathrm{GeV}\,\frac{L}{1300~\mathrm{km}}
	\,\frac{\abs{\dm_{31}}}{2.5\times 10^{-3}~\mathrm{eV}^2}\,.
\end{equation*}

\acknowledgments

The  authors  would  like  to  acknowledge  fruitful  discussions  with  
Francisco  Botella,  Anselmo  Cervera,  Sergio  Palomares and Michel Sorel.
This research has been supported by MINECO Project FPA 2017-84543-P, 
Generalitat Valenciana Project GV PROMETEO 2017-033 and 
Severo Ochoa Excellence Centre Project SEV 2014-0398. 
A.S. acknowledges the MECD support through the FPU14/04678 grant.


\begin{thebibliography}{99}

%%%%%%%%%%%%%%%%%%%%%%%%
%     Introduction
%%%%%%%%%%%%%%%%%%%%%%%%

\bibitem{baryogenesis}
M. Fukugita and T. Yanagida,
\emph{Baryogenesis Without Grand Unification}, 
\emph{Phys.Lett.} {\bf B174} (1986) 45.
%\href{https://doi.org/10.1016/0370-2693(86)91126-3}{Phys.Lett. {\bf B174}, 45 (1986)}.

\bibitem{HK}
K. Abe et al. [Hyper-Kamiokande Proto-Collaboration],
\emph{Hyper-Kamiokande design report},
KEK-Preprint-2016-21, ICRR-Report-701-2016-1.
%\href{http://www.hyperk.org/?p=215}{KEK-Preprint-2016-21, ICRR-Report-701-2016-1}.

\bibitem{DUNE}
R. Acciarri et al. [DUNE Collaboration],
\emph{Long-Baseline Neutrino Facility (LBNF) and Deep Underground Neutrino Experiment (DUNE) Conceptual Design Report Volume 2: The Physics Program for DUNE and LBNF},
FERMILAB-DESIGN-2016-02
[arXiv:1512.06148 [physics.ins-det]].

\bibitem{MSW-W}
L. Wolfenstein,
\emph{Neutrino oscillations in matter},
\emph{Phys.Rev.D} {\bf 17} (1978) 2369.
%\href{https://doi.org/10.1103/PhysRevD.17.2369}{Phys.Rev.D {\bf 17}, 2369 (1978)}.

\bibitem{MSW-MS}
S.P. Mikheyev and A.Yu. Smirnov,
\emph{Resonance enhancement of oscillations in matter and solar neutrino spectroscopy},
\emph{Sov.J.Nucl.Phys.} {\bf 42} (1985) 913.

\bibitem{disentanglingPRL}
J. Bernabeu and A. Segarra,
\emph{Disentangling genuine from matter-induced CP violation in neutrino oscillations},
[arXiv:1806.07694 [hep-ph]].



%%%%%%%%%%%%%%%%%%%%%%%%
%     Teorema
%%%%%%%%%%%%%%%%%%%%%%%%

\bibitem{Matter-Hamiltonian-Barger}
V. Barger, K. Whisnant, S. Pakvasa and R.J.N. Phillips, 
\emph{Matter effects on three-neutrino oscillations},
\emph{Phys.Rev.D} {\bf 22} (1980) 2718.
%\href{https://doi.org/10.1103/PhysRevD.22.2718}{Phys.Rev.D {\bf 22}, 2718 (1980)}.

\bibitem{Matter-Hamiltonian-Kuo}
T.K. Kuo and J. Pantaleone, 
\emph{Neutrino oscillations in matter},
\emph{Rev.Mod.Phys.} {\bf 61} (1989) 937.
%\href{https://doi.org/10.1103/RevModPhys.61.937}{Rev.Mod.Phys. {\bf 61}, 937 (1989)}.

\bibitem{Matter-Hamiltonian-Zaglauer}
H.W. Zaglauer and K.H. Schwarzer, 
\emph{The mixing angles in matter for three generations of neutrinos and the
MSW mechanism},
\emph{Z.Phys.} {\bf C40} (1988) 273.
%\href{https://doi.org/10.1007/BF01555889}{Z.Phys. {\bf C40}, 273 (1988)}.

\bibitem{Matter-Hamiltonian-Krastev}
P. Krastev, 
\emph{Searching for the MSW effect with neutrino beams from next
generation accelerators},
\emph{Nuovo Cim.} {\bf A103} (1990) 361.
%\href{https://doi.org/10.1007/BF02790019}{Nuovo Cim. {\bf A103}, 361 (1990)}.

\bibitem{Matter-Hamiltonian-Parke}
R.H. Bernstein and S.J. Parke, 
\emph{Terrestrial long-baseline neutrino-oscillation experiments},
\emph{Phys.Rev.D} {\bf 44} (1991) 2069.
%\href{https://doi.org/10.1103/PhysRevD.44.2069}{Phys.Rev.D {\bf 44}, 2069 (1991)}.




%%%%%%%%%%%%%%%%%%%%%%%%
%     Desarrollos
%%%%%%%%%%%%%%%%%%%%%%%%

\bibitem{petcov}
S.T. Petcov and Y.-L. Zhou,
\emph{On Neutrino Mixing in Matter and CP and T Violation Effects in Neutrino Oscillations},
[arXiv:1806.09112 [hep-ph]].

\bibitem{shunCPT}
T. Ohlsson and S. Zhou,
\emph{Extrinsic and Intrinsic CPT Asymmetries in Neutrino Oscillations},
\emph{Nucl.Phys.} {\bf B893} (2015) 482
%\href{https://doi.org/10.1016/j.nuclphysb.2015.02.015}{Nucl.Phys. {\bf B893}, 482 (2015)}.
[arXiv:1408.4722 [hep-ph]].

\bibitem{xing}
Z-z. Xing,
\emph{New Formulation of Matter Effects on Neutrino Mixing and CP Violation},
\emph{Phys.Lett.} {\bf B487} (2000) 327
%\href{https://doi.org/10.1016/S0370-2693(00)00832-7}{Phys.Lett. {\bf B487}, 327 (2000)}.
[arXiv:hep-ph/0002246].

\bibitem{cervera}
A. Cervera, A. Donini, M.B. Gavela, J.J. Gomez Cadenas, P. Hernandez, O. Mena and S. Rigolin,
\emph{Golden measurements at a neutrino factory},
\emph{Nucl.Phys.} {\bf B593} (2001) 731
%\href{https://doi.org/10.1016/S0550-3213(00)00221-2}{Nucl.Phys. {\bf B593}, 731 (2001)}.
[arXiv:hep-ph/0002108].

\bibitem{parke}
P.B. Denton, H. Minakata and S.J. Parke,
\emph{Compact perturbative expressions for neutrino oscillations in matter},
\emph{JHEP} {\bf 1606} (2016) 051
%\href{https://doi.org/10.1007/JHEP06(2016)051}{JHEP {\bf 1606}, 051 (2016)}.
[arXiv:1604.08167 [hep-ph]].

\bibitem{ioannisian}
Ara Ioannisian and S. Pokorski,
\emph{Three Neutrino Oscillations in Matter},
\emph{Phys.Lett.} {\bf B782} (2018) 641
%\href{https://doi.org/10.1016/j.physletb.2018.06.001}{Phys.Lett. {\bf B782}, 641 (2018)}.
[arXiv:1801.10488 [hep-ph]].

\bibitem{xing_pert}
Z-z. Xing and J-y. Zhu,
\emph{Analytical approximations for matter effects on CP violation in the
accelerator-based neutrino oscillations with $E \lesssim 1$ GeV},
\emph{JHEP} {\bf 1607} (2016) 011
%\href{https://doi.org/10.1007/JHEP07(2016)011}{JHEP {\bf 1607}, 011 (2016)}.
[arXiv:1603.02002 [hep-ph]].

\bibitem{mariam}
P.F. de Salas, D.V. Forero, C.A. Ternes, M. Tortola and F.W.F. Valle,
\emph{Status of neutrino oscillations 2018: 3$\sigma$ hint for normal mass ordering and improved CP sensitivity},
\emph{Phys.Lett.} {\bf B782} (2018) 633
%\href{https://doi.org/10.1016/j.physletb.2018.06.019}{Phys.Lett. {\bf B782}, 633 (2018)}.
[arXiv:1708.01186v2 [hep-ph]]

\bibitem{matter0th}
M.C. Banuls, G. Barenboim and J. Bernabeu,
\emph{Medium effects for terrestrial and atmospheric neutrino oscillations},
\emph{Phys.Lett.} {\bf B513} (2001) 391
%\href{https://doi.org/10.1016/S0370-2693(01)00723-7}{Phys.Lett. {\bf B513}, 391 (2001)}.
[arXiv:hep-ph/0102184].

\bibitem{a=3E}
I. Mocioiu and R. Shrock, 
\emph{Matter effects on neutrino oscillations in long baseline experiments},
\emph{Phys.Rev.} {\bf D62} (2000) 053017
%\href{https://doi.org/10.1103/PhysRevD.62.053017}{Phys.Rev. {\bf D62}, 053017 (2000)}.
[arXiv:hep-ph/0002149].


%%%%%%%%%%%%%%%%%%%%%%%%
%     Genuine CPV
%%%%%%%%%%%%%%%%%%%%%%%%

\bibitem{CP-condition}
J. Bernabeu, G.C. Branco and M. Gronau,
\emph{CP Restrictions on Quark Mass Matrices},
\emph{Phys.Lett.} {\bf B169} (1986) 243.
%\href{https://doi.org/10.1016/0370-2693(86)90659-3}{Phys.Lett. {\bf B169}, 243 (1986)}.

\bibitem{Harrison-Scott}
P.F. Harrison and W.G. Scott,
\emph{CP and T Violation in Neutrino Oscillations and Invariance of Jarlskog's
Determinant to Matter Effects},
\emph{Phys.Lett.} {\bf B476} (2000) 349
%\href{https://doi.org/10.1016/S0370-2693(00)00153-2}{Phys.Lett. {\bf B476}, 349 (2000)}.
[arXiv:hep-ph/9912435].

\bibitem{Jarlskog}
C. Jarlskog,
\emph{A basis independent formulation of the connection between quark mass
matrices, CP violation and experiment},
\emph{Z.Phys.} {\bf C29} (1985) 491.
%\href{https://doi.org/10.1007/BF01565198}{Z.Phys. {\bf C29}, 491 (1985)}.


%%%%%%%%%%%%%%%%%%%%%%%%
%     Hierarchy
%%%%%%%%%%%%%%%%%%%%%%%%

\bibitem{shao-feng}
S.-F. Ge, K. Hagiwara and C. Rott,
\emph{A Novel Approach to Study Atmospheric Neutrino Oscillation},
\emph{JHEP} {\bf 1406} (2014) 150
%\href{https://doi.org/10.1007/JHEP06(2014)150}{JHEP {\bf 1406}, 150 (2014)}.
[arXiv:1309.3176 [hep-ph]]

\bibitem{invarRe_kimura}
K. Kimura, A. Takamura and H. Yokomakura,
\emph{Exact formula of probability and CP violation for neutrino oscillations
in matter},
\emph{Phys.Lett.} {\bf B537} (2002) 86
%\href{https://doi.org/10.1016/S0370-2693(02)01907-X}{Phys.Lett. {\bf B537}, 86 (2002)}.
[arXiv:hep-ph/0203099].

\bibitem{invarRe_scott}
P.F. Harrison, W.G. Scott and T.J. Weiler,
\emph{Exact matter-covariant formulation of neutrino oscillation probabilities},
\emph{Phys.Lett.} {\bf B565} (2003) 159
%\href{https://doi.org/10.1016/S0370-2693(03)00749-4}{Phys.Lett. {\bf B565}, 159 (2003)}.
[arXiv:hep-ph/0305175].


%%%%%%%%%%%%%%%%%%%%%%%%
%     Energy bin
%%%%%%%%%%%%%%%%%%%%%%%%

\bibitem{fdez-mtnez}
V. De Romeri, E. Fernandez-Martinez and M. Sorel,
\emph{Neutrino oscillations at DUNE with improved energy reconstruction},
\emph{JHEP} {\bf 1609} (2016) 030
%\href{https://doi.org/10.1007/JHEP09(2016)030}{JHEP {\bf 1609} (2016) 030}
[arXiv:1607.00293 [hep-ph]]




\end{thebibliography}
\end{document}